\documentclass[lettersize,journal]{IEEEtran}

\usepackage{cite}
\usepackage{amsmath,amssymb,amsfonts}
\usepackage{algorithm2e}
\SetKwRepeat{Do}{do}{while}%
\RestyleAlgo{ruled}

\usepackage{balance}
\usepackage{graphicx}
\usepackage{textcomp}
\usepackage{xcolor}
\usepackage{caption}
\usepackage{subcaption}
\usepackage{makecell}
\usepackage{multirow}
\usepackage{enumitem}

\usepackage{caption}
\usepackage{colortbl}
\usepackage{color}
\definecolor{light_grey}{rgb}{0.8,0.8,0.8}
\usepackage{dblfloatfix}

\usepackage{comment}

\DeclareMathOperator*{\argmax}{arg\,max}
\DeclareMathOperator*{\argmin}{arg\,min}

\IEEEoverridecommandlockouts
\usepackage{tikz}
\usepackage{textcomp}
\usepackage[hidelinks]{hyperref}
\usepackage{lipsum}
\newcommand\copyrighttext{%
  \footnotesize \textcopyright 2025 IEEE. Personal use of this material is permitted.  Permission from IEEE must be obtained for all other uses, in any current or future  media, including reprinting/republishing this material for advertising or promotional  purposes, creating new collective works, for resale or redistribution to servers or  lists, or reuse of any copyrighted component of this work in other works.
  DOI: \href{<http://tex.stackexchange.com>}{10.1109/TWC.2025.3547648}
  }
\newcommand\copyrightnotice{%
\begin{tikzpicture}[remember picture,overlay]
\node[anchor=south,yshift=10pt] at (current page.south) {\fbox{\parbox{\dimexpr\textwidth-\fboxsep-\fboxrule\relax}{\copyrighttext}}};
\end{tikzpicture}%
}

\begin{document}

\title{Resource allocation exploiting reflective surfaces to minimize the outage probability in VLC}

\author{Borja Genoves Guzman,~\IEEEmembership{Senior Member,~IEEE,}
Maximo Morales Cespedes,~\IEEEmembership{Member,~IEEE,}
Victor P. Gil Jimenez,~\IEEEmembership{Senior Member,~IEEE,}
Ana Garcia Armada,~\IEEEmembership{Fellow,~IEEE,}
Maïté Brandt-Pearce,~\IEEEmembership{Fellow,~IEEE,}
\thanks{This work has been partially funded by project PID2023-147305OB-C31 (SOFIA-AIR) by MCIN/ AEI/10.13039/501100011033/ERDF, UE, and by project TED2021-129869B-I00 funded by MICIU/AEI/10.13039/501100011033 and Unión Europea NextGenerationEU/PRTR. Borja Genoves Guzman has received funding from the European Union under the Marie Skłodowska-Curie grant agreement No 101061853, and M. Morales Cespedes from Ramón y Cajal grant RYC2022-036053-I, MICIU/AEI/10.13039/501100011033 and FSE+.}
\thanks{Borja Genoves Guzman, Maximo Morales Cespedes, Victor P. Gil Jimenez and Ana Garcia Armada are with Universidad Carlos III de Madrid, Depto. Teoría de la Señal y Comunicaciones, Av. de la Universidad 30, 28911 Leganés, Madrid, Spain. Maïté Brandt-Pearce is with University of Virginia, Dept. Electrical and Computer Engineering, Charlottesville, VA 22904 USA. E-mails: bgenoves@ing.uc3m.es, maximo@tsc.uc3m.es, vgil@tsc.uc3m.es, agarcia@tsc.uc3m.es, mb-p@virginia.edu}
\vspace{-8mm}
}




\maketitle

\copyrightnotice 
\vspace{-\baselineskip}

\begin{abstract}
Visible light communication (VLC) is a technology that complements radio frequency (RF) to fulfill the ever-increasing demand for wireless data traffic. The ubiquity of light-emitting diodes (LEDs), exploited as transmitters, increases the VLC market penetration and positions it as one of the most promising technologies to alleviate the spectrum scarcity of RF. However, VLC deployment is hindered by blockage causing connectivity outages in the presence of obstacles. Recently, optical reconfigurable intelligent surfaces (ORISs) have been considered to mitigate this problem. While prior works exploit ORISs for data or secrecy rate maximization, this paper studies the optimal placement of mirrors and ORISs, and the LED power allocation, for jointly minimizing the outage probability while keeping the lighting standards. 
We describe an optimal outage minimization framework called \textsc{JointMinOut} and present solvable heuristics. 
We provide extensive numerical results and show that the use of ORISs may reduce the outage probability by up to 67\% with respect to a no-mirror scenario and provide a gain of hundreds of kbit/J in optical energy efficiency with respect to the presented benchmark.
\end{abstract}

\begin{IEEEkeywords}
Line-of-sight (LoS) link blockage, mirrors, optimal placement, outage probability, optical reconfigurable intelligent surfaces (ORIS), visible light communication (VLC).
\end{IEEEkeywords}

\section{Introduction}\label{Intro}

\IEEEPARstart{V}{isible} light communication (VLC) has recently emerged as a technology to complement traditional radio-frequency (RF) in the mission of satisfying the ever-increasing demand for wireless data traffic.  RF resources are becoming crowded and fragmented, and the research community is looking for
alternatives to meet the requirements beyond 5G. VLC presents unique characteristics such as unused and unlicensed bands, off-the-shelf elements, higher security in the physical layer due to a better signal containment, and a ubiquitous light-emitting diode (LED)-based lighting infrastructure to exploit for communications~\cite{LEDinUSA}. 
The first VLC-related standards were published by ITU-T~\cite{G9991} and IEEE\cite{IEEE8021513}\cite{IEEE802157}, but VLC industrialization efforts have only been addressed in the recently published IEEE 802.11bb standard~\cite{IEEE80211bb}, where VLC has been brought into the WiFi ecosystem to push its mass market adoption. 
Even so, VLC still presents severe issues that prevent it from a massive adoption, among which are link blockages causing connectivity outages in the presence of obstacles that make VLC unreliable~\cite{LoSblockageImpact}. Prior works show that this issue can be partially solved by cooperative techniques~\cite{LightsAndShadows} or relaying schemes~\cite{ReflectionBasedRelayVLC}, leveraging the high reuse factor of resources when operating in very small cells (usually referred to as atto-cells) created by VLC when it is integrated as a cellular network~\cite{WhatIsLiFi}. In this work we exploit reflective surfaces and optimize resource allocation to minimize the outage probability in VLC.

Reconfigurable intelligent surfaces (RISs) have been studied in RF for short distances employing millimeter and sub-millimeter waves~\cite{IRS_mmWave, IRS_THz}, and for long distances employing lower frequencies~\cite{RIS_RF_longDist}. RIS can be managed dynamically, and it 
is foreseen to be one fundamental pillar of 6G networks~\cite{RIS6G, RIS6G2}. It has emerged as a solution to the skip-zone problem in RF~\cite{RISRF}. 
However, although  link blockage is one of the main drawbacks of VLC, we can only find a few studies of optical RIS (ORIS) in the VLC literature~\cite{RISVLC}.  Unlike RIS systems in RF, ORIS-assisted VLC systems do not suffer from small scale fading as the system uses intensity modulation with direct detection (IM/DD), and the detection area is very large compared to the optical wavelength. That is, while RIS in RF must pay special attention to the phase of each element in the RIS matrix to achieve the largest gain, ORIS must focus on forwarding the impinging light power to the right direction~\cite{FD_IRS_VLC}. Therefore, RIS-assisted RF studies cannot be directly adapted to VLC systems. 

ORISs have been proposed to engineer non-line-of-sight (NLoS) paths in VLC and then to enhance its wireless communication capability. Recent research in ORIS-assisted VLC includes indoor~\cite{JointResourceMirrors, MirrorVLC}, free space~\cite{FreeSpaceOptics_RIS, FreeSpaceOptics_RIS2}, vehicular~\cite{RIS_vlc_v2v} and unmanned aerial vehicle~\cite{RIS_vlc_uav} applications.  Published works have addressed channel modeling in ORIS-aided VLC systems in the time-domain~\cite{TD_IRS_VLC} and in the frequency-domain~\cite{FD_IRS_VLC}. Two types of ORIS have been proposed in the literature: metasurfaces and mirrors. The former is based on the meta-atom geometry and can actuate over the wavelength, amplitude, and polarization of the light signal, among others. The latter is based on Snell's law to modify the reflection angle by changing the mirror orientation. Although metasurfaces show a great potential as ORIS in VLC, deployed on walls and to improve the receiver performance~\cite{DRIS_OWC, DRIS_OWC2, ORIS_rx, ORIS_rx2}, few products can be found on the market. The excellent reflection performance of mirrors, together with the recent progress in micro-electromechanical systems (MEMS) used to manage mirror orientation, position mirrors as an imminent ORIS material of choice for VLC~\cite{MirrorVSMetasurface}. 

Few works invoking mirrors as ORIS in VLC systems can be found in the literature:~\cite{IRSIndoorVLC} and~\cite{SumRateMirrors} proposed a rate maximization problem to determine the optimal orientation of the mirror array elements and the association between light source and mirror to strengthen the NLoS link; the authors in~\cite{JointResourceMirrors} maximized the overall spectral efficiency by optimizing the user association and power allocation; other recent works studied the secrecy rate~\cite{SecrecyRateVLC1, SecrecyRateVLC2}; and~\cite{MirrorVLC} proposed an optimal mirror placement to maximize the illumination uniformity, and then associated users and LEDs to maximize the minimum signal-to-interference-plus-noise power ratio (SINR).

In this paper, we refer to {\em mirrors} when they are static and flat against the wall, and {\em ORISs} when installed into MEMS structures to provide them with a mobile orientation. We evaluate both options to minimize the outage probability in an indoor VLC scenario by jointly optimizing the optical power allocated per LED and the number and location of mirrors (or ORIS). As this is a difficult problem, we target the theoretical limit of the outage probability in our study by limiting our analysis to a single user. We consider any user location and orientation by uniformly distributing a single user and analyzing the frequency of occurrence for each optimal mirror placement. Note that this is key in the design of a mirror- or ORIS-aided VLC scenario, as the installation of mirrors and ORIS elements is fixed; changing the mirrors and ORIS elements placement for each case (user position and orientation) is not practical. Thus, the mirrors or ORIS elements placement should be optimized such that their contribution is maximized. We name our optimization problem \textsc{JointMinOut}. 

Despite the fact that outage is one of the biggest problems in VLC, this is, to the authors' knowledge, the first work that considers reflective surfaces to minimize the outage probability. We build upon our prior work where we simplified the scenario to a single serving LED for each user and considered only those users with a line-of-sight (LoS)-link blocked~\cite{Globecom2023}. Besides, in~\cite{Globecom2023}, a diffuse channel path loss was generalized and illumination constraints were not considered. 
The main contributions of this paper \mbox{are summarized as follows:}
\begin{itemize}[leftmargin=*]
\item We introduce three possible reflecting elements in a room: wall, mirror, and ORIS, and we explain the fundamentals of their diffuse and specular reflections. In our system, we do not have a pre-established mirror/ORIS structure, but we optimize the number of mirror/ORIS elements and their location. We show that diffuse wall reflections cannot be ignored, as done in prior ORIS-assisted VLC works. Besides, we show that NLoS contributions from mirrors/ORISs can be even larger than the LoS.
\item We compare the coverage that mirror and ORIS approaches can achieve, and we derive equations to evaluate such coverage in both cases. We show that the furthest user location to be supported by a mirror depends on the LED position and the field of view (FoV) semi-angle of the receiver, whereas it only depends on the FoV semi-angle in the ORIS case.
\item We formulate our \textsc{JointMinOut} optimization problem to minimize the outage probability while minimizing one of the two available resources, the number of mirrors (or ORISs) and the total optical power allocated among LEDs. Unlike~\cite{Globecom2023}, we consider a scenario where all LEDs distributed in the room may contribute to the same user. We introduce communication and illumination constraints and re-formulate them to be solved by common convex optimization (CVX) tools. Then, we propose two single-objective functions leading to two solution approaches.
\item The resulting problems are NP-complete, and we propose an alternating optimization heuristic algorithm to solve them. The results are compared with a non-alternating heuristic algorithm as a benchmark and a no-mirror scenario.
\item We offer some insights on deploying the mirrors or ORISs to minimize the outage probability. Simulation results show that mirrors or ORISs, if optimally located along the wall, can reduce the outage probability up to 67\% compared to a no-mirror scenario, as well as providing considerable optical energy efficiency gains while keeping a low complexity profile. 
\end{itemize}

The structure of the paper is as follows. Section\,\ref{SystemModel} introduces the system model, including LoS and NLoS channel gains for every reflector type, a comparison between mirror and ORIS performance, and the figures of merit used in this paper. Section\,\ref{OptimizationProblem} formulates the optimization problem. The algorithms proposed to solve the optimization problem are presented in Section\,\ref{AlgorithmsProposed}, and detailed results and discussions are contained in Section\,\ref{ResultsAndDiscussion}. Finally, conclusions are drawn in Section\,\ref{sec:Conclusion}.


\vspace{-2mm}
\section{System model}\label{SystemModel}
The considered indoor VLC scenario has $L$ LEDs distributed in the room, denoted by $l=\{0,\cdots,L{-}1\}$. All LEDs may contribute to the same user, i.e., they cooperate to transmit the same information to the user. LEDs are modeled as point sources due to their small dimensions, and detectors are small enough to not perceive irradiance variations along their surface. Then we can also assume single-point detectors. In this paper, we study a single user whose location is uniformly distributed inside the room. Multiple users could be easily served invoking multiple access techniques such as time-division multiple access. We assume that the user equipment and LED are looking upwards and downwards, respectively. The communication performance relies on LoS and NLoS links defined as follows. 


\begin{figure}[t]
     \centering
     \begin{subfigure}[t]{0.37\columnwidth}
         \centering
         \includegraphics[width=\textwidth]{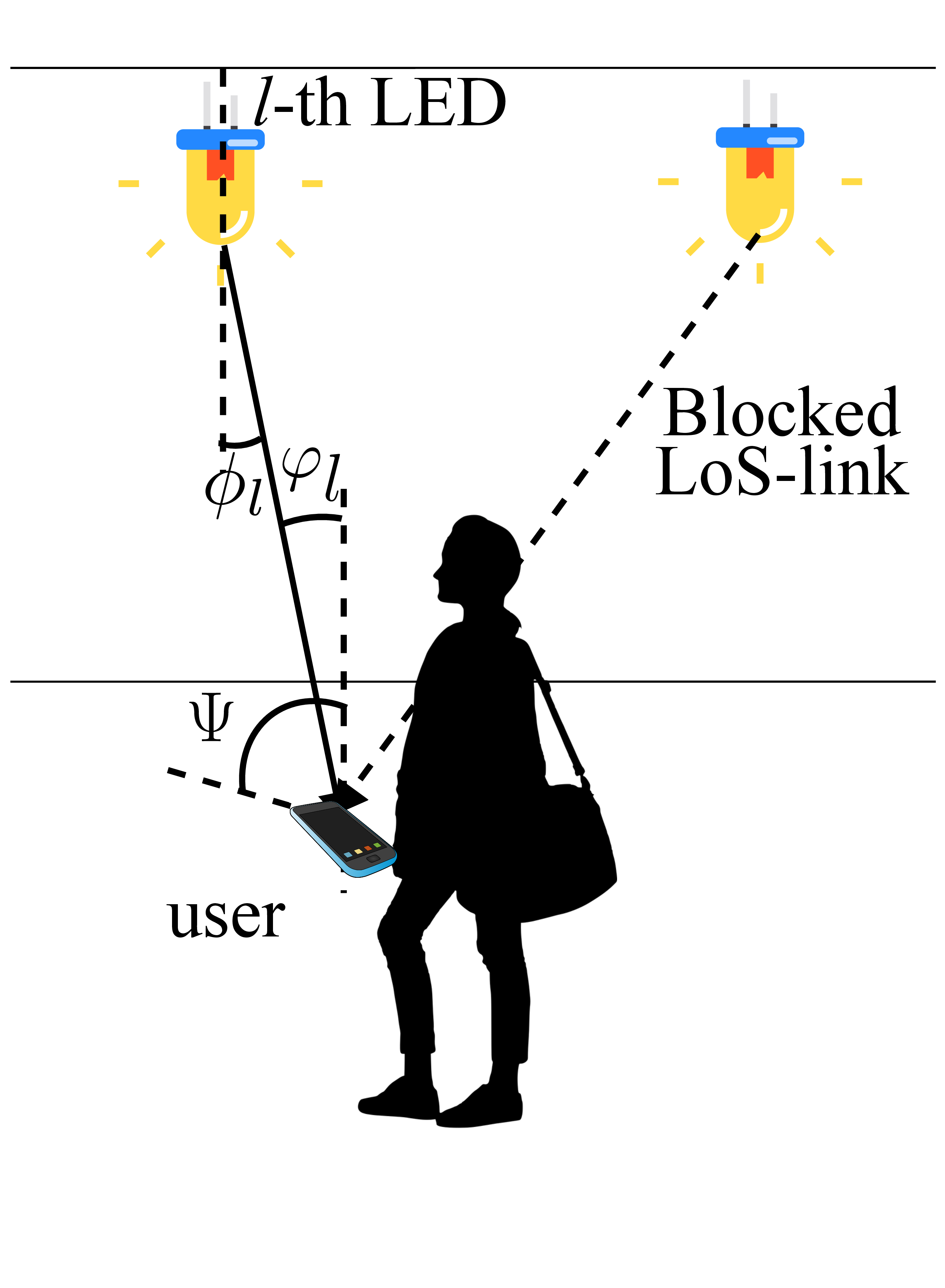}
         \vspace{-2mm}\caption{LoS}
         \label{fig:ScenarioLoS}
     \end{subfigure}
     \hfill
     \begin{subfigure}[t]{0.57\columnwidth}
         \centering
         \includegraphics[width=\textwidth]{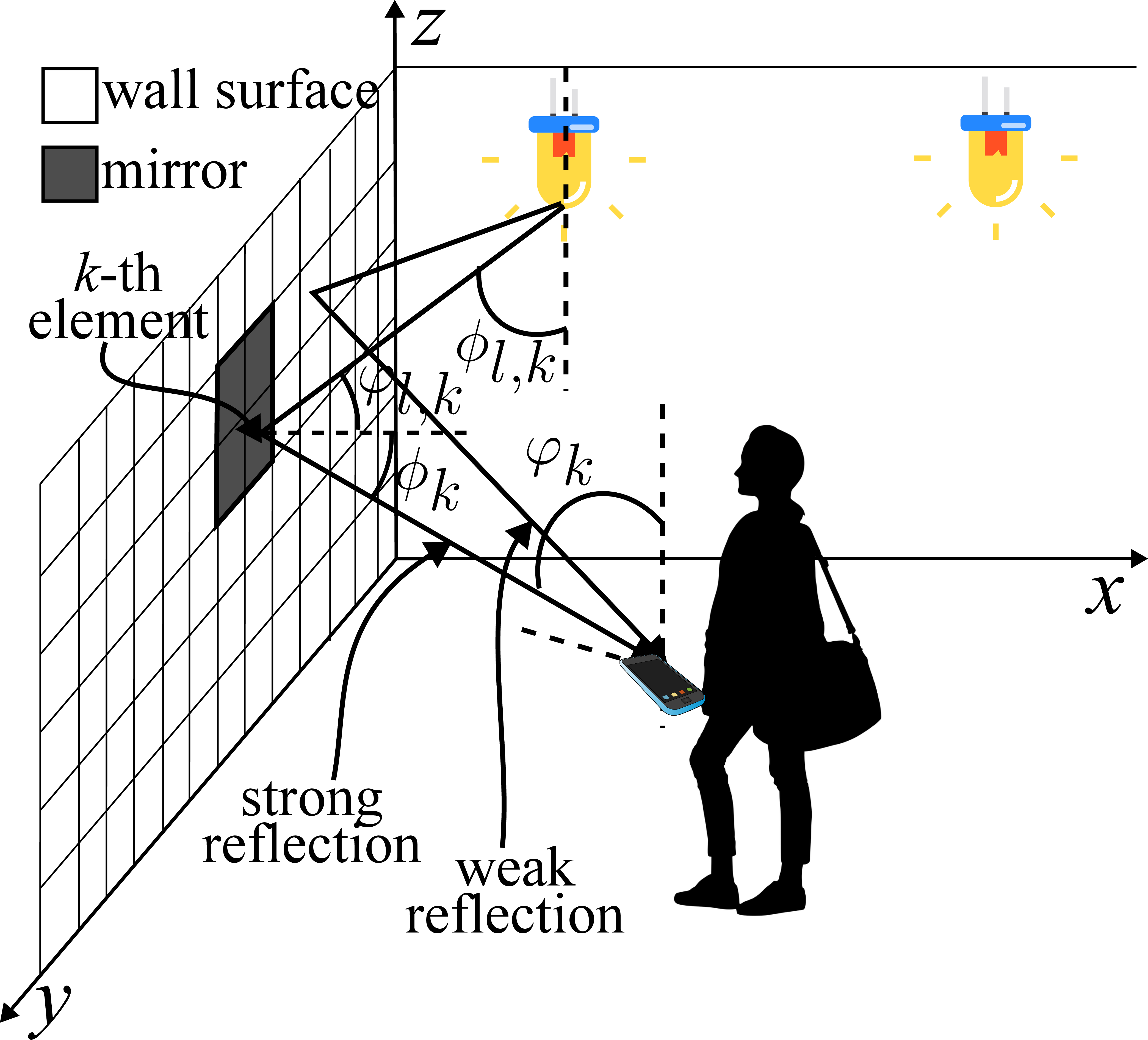}
         \vspace{-2mm}\caption{NLoS}
         \label{fig:ScenarioNLoS}
     \end{subfigure}
     \vspace{-2mm}
        \caption{3D illustration of LoS and NLoS propagation. Light sources are located on a horizontal plane on the ceiling of the room.}        
        \label{fig:SystemModel}
        \vspace{-4mm}
\end{figure}

\subsection{LoS channel gain}
The LoS channel gain from LED $l$ to the user follows a Lambertian emission model as~\cite{VLCLoSChannel}
\vspace{-1mm}
\begin{equation}
\label{eq:ChannelModelLoS}
H_{l}^{{\rm{LoS}}} =
\begin {cases}
\frac{{\left( {m + 1} \right) \cdot {A_{{\rm{PD}}}}}}{{2\pi d_{l}^2}}{{\cos }^m}\left( {{\phi _{l}}} \right)\cos \left( {{\varphi _{l}}} \right) & 0 \le {\varphi _{l}} \le {\Psi}\\ 
0 & \rm{otherwise},
\end{cases}
\vspace{-2mm}
\end{equation}
where $m=-1/\log_2\left(\cos\left(\phi_{1/2}\right)\right)$ is the Lambertian index of the LED that models the radiation pattern defined by its half-power semi-angle $\phi_{1/2}$. The parameter $A_{{\rm PD}}$ stands for the active photodetector (PD) area, $d_{l}$ is the Euclidean distance between LED $l$ and the user, and $\phi_{l}$ and $\varphi_{l}$ are the irradiance and incidence angles, respectively, as represented in Fig.\,\ref{fig:ScenarioLoS}. The FoV semi-angle of the PD is denoted by $\Psi$.

\begin{figure}[t]
     \centering
     \begin{subfigure}[b]{0.32\columnwidth}
         \centering
         \includegraphics[width=\textwidth]{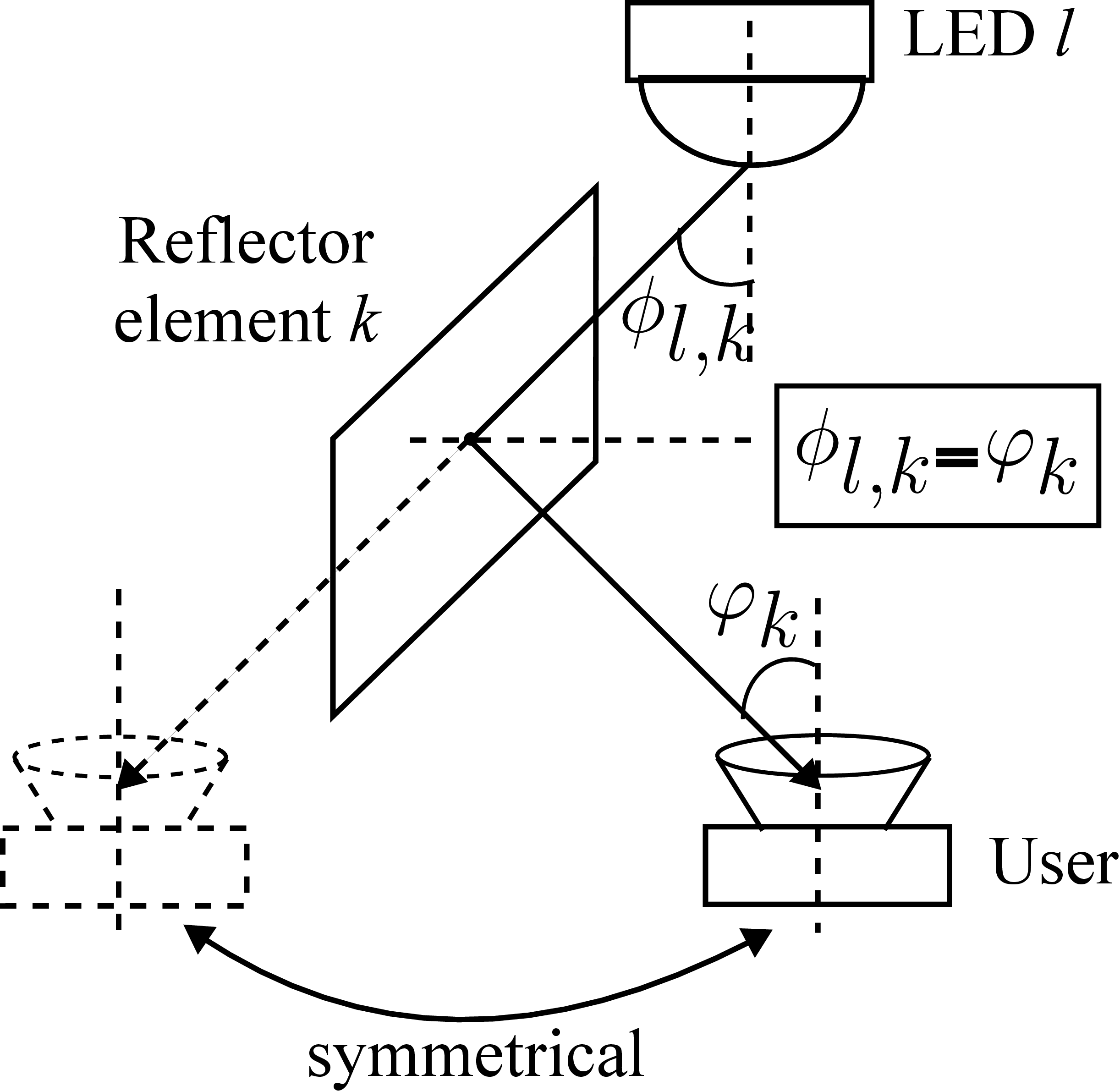}
         \caption{}
         \label{fig:Mirror}
     \end{subfigure}
     \hfill
     \begin{subfigure}[b]{0.32\columnwidth}
         \centering
         \includegraphics[width=\textwidth]{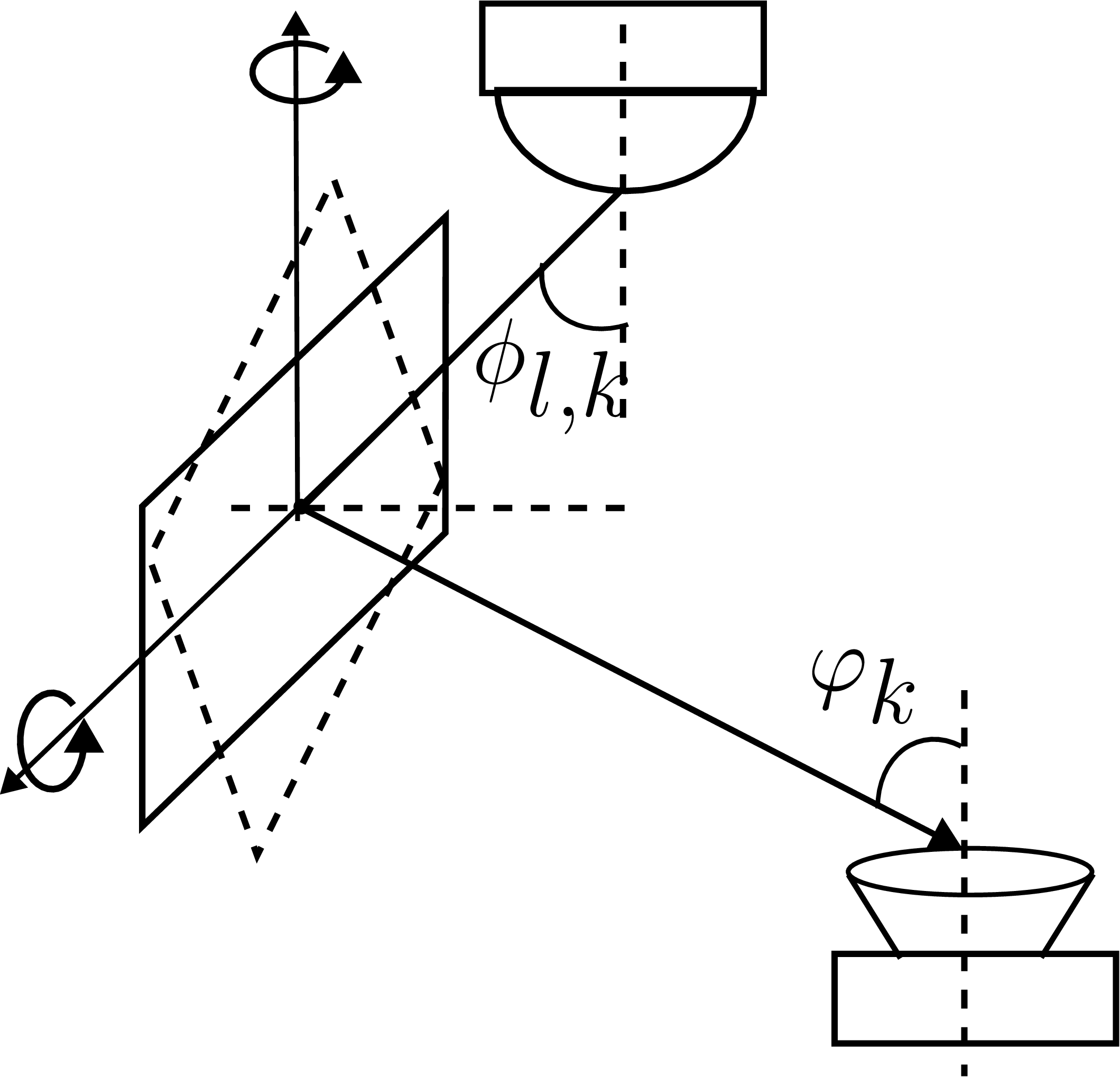}
         \caption{}
         \label{fig:RIS}
     \end{subfigure}
          \hfill
  \begin{subfigure}[b]{0.32\columnwidth}
         \centering
         \includegraphics[width=\textwidth]{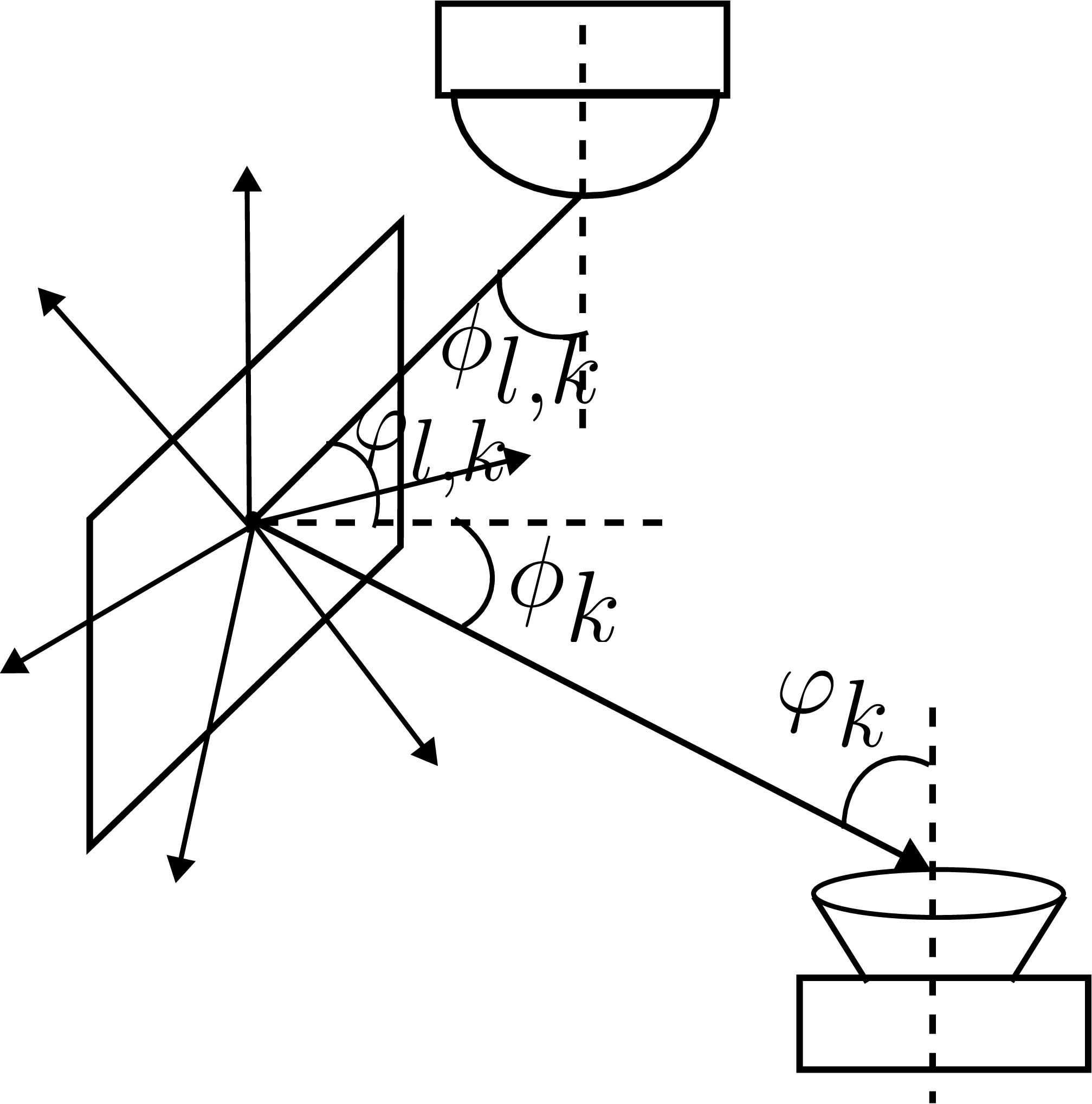}
         \caption{}
         \label{fig:Wall}
     \end{subfigure}
     \vspace{-2mm}
        \caption{Three types of reflecting surfaces considered: (a) mirror (specular channel), (b) mobile mirror = ORIS (specular channel), and (c) wall (diffuse channel).}        
        \label{fig:ReflectingSurfaces}
        \vspace{-5mm}
\end{figure}

\vspace{-4mm}
\subsection{NLoS channel gain}
\vspace{-1mm}
The NLoS channel in VLC can be composed of specular and diffuse reflections. Specular reflections refer to those that follow a unique direction; the main losses occur due to the medium absorption. The primary specular reflectors in VLC are mirrors. Differently, diffuse reflections occur when the reflector does not present a homogeneous surface and, when light impinges onto it, the light scatters in multiple directions, producing a considerable power loss when targeting one specific receiver. 

In this paper, we consider that each wall is divided into a grid of $K_{\rm{y}}\times K_{\rm{z}}\,=\,K$ elements, denoted by $k=\{0,\cdots,K{-}1\}$ as represented in Fig.\,\ref{fig:ScenarioNLoS}. Each element is either selected as a mirror surface or as a standard wall material. Within the mirror category, this can be: (1) installed in a MEMS structure 
to create a mobile mirror, which we here refer to as an ORIS; or (2) installed following the wall orientation without mobility,  which we simply refer to as a mirror. These two mirror surfaces, together with the wall surface, are depicted in Fig.\,\ref{fig:ReflectingSurfaces}. We assume that blockage between reflectors is negligible. The NLoS channel gain produced by each is detailed in the following:
\subsubsection{Mirror reflection (specular)} The planar surface reflects the incident light with the same angle, following Snell's law of reflection~\cite{Snell}, due to its static and flat-against-the-wall placement. Since the mirror orientation is fixed, the irradiance angle from LED $l$ to the mirror reflecting point $k$ ($\phi _{l,k}$) equals the incident angle from the reflecting point $k$ to the PD ($\varphi_{k}$). The NLoS channel gain produced by a mirror can be modeled as~\cite{MirrorVSMetasurface}
\vspace{-2mm}
\begin{equation}
\label{eq:ChannelModelNLoSmirror}
\begin{split}
&H_{l,k}^{{\rm{mirror}}} =
\begin {cases}
{\hat{r} {\cdot} {\frac{{\left( {m + 1} \right) {\cdot} {A_{{\rm{PD}}}}}}{{2\pi {{\left( {{d_{l,k}} + {d_{k}}} \right)}^2}}}{{\cos }^{m+1}}{\left( {{\phi _{l,k}}} \right)}}} & \hspace{0.8cm}{ 0 {\le} {\varphi _{k}} {\le} {\Psi}} \text{ \&} \\ & \hspace{-3cm}\varphi_{k}=\phi_{l,k}\text{ at some point in element $k$,} \\ 
0 & \hspace{1.1cm}\text{otherwise},
\end{cases}
\end{split}
\vspace{-2mm}
\end{equation}
where $\hat{r}$ stands for the reflection coefficient of a specular surface (e.g. mirror or ORIS), $d_{l,k}$ and $d_{k}$ are the Euclidean distance from LED $l$ to reflector element $k$, and from reflector element $k$ to the user, respectively. 
Note that a specular reflection is equivalent to considering a user positioned at the image point, where the total distance is the sum of the two distances in the path LED$\rightarrow$mirror$\rightarrow$user, as represented in Fig.\,\ref{fig:Mirror}.
\subsubsection{Mobile mirror (ORIS) reflection (specular)} In this case the channel model can be formulated as~\cite{MirrorVSMetasurface}
\begin{equation}
\label{eq:ChannelModelNLoSRIS}
H_{l,k}^{{\rm{ORIS}}} =
\begin {cases}
\hspace{-1mm}{{\hat{r}} {\cdot} {\frac{{\left( {m + 1} \right) \cdot {A_{{\rm{PD}}}}}}{{2\pi {{\left( {{d_{l,k}} + {d_{k}}} \right)}^2}}}{{\cos }^m}{\left( {{\phi _{l,k}}} \right)}{\cos} \left( {{\varphi _{k}}} \right)}} & { 0 {\le} {\varphi _{k}} {\le} {\Psi}} \\
\hspace{-1mm}0 & \rm{otherwise}. \\
\end{cases}
\end{equation}
Due to its installation into a MEMS structure, the ORIS orientation mobility controls the reflected path direction, as shown in Fig.\,\ref{fig:RIS}, so that the impinging light power is forwarded to the direction where the user is located. 
\subsubsection{Wall reflection (diffuse)}As Fig.\,\ref{fig:Wall} shows, the light impinging onto a wall surface is reflected in multiple directions. The channel gain can be modeled as~\cite{VLCLoSModel}
\vspace{-2mm}
\begin{equation}
\label{eq:ChannelModelNLoSdiff}
\hspace{-5mm}
H_{l,k}^{{\rm{wall}}} {=}\hspace{-1mm}
\begin {cases}
{{\tilde{r}} {\cdot} \frac{{\left( {m + 1} \right)  {A_{{\rm{PD}}}}}}{{2{\pi}d_{l,k}^2d_{k}^2}}\hspace{-1mm}{A_k}{{\cos }^m}{\left( {{\phi _{l,k}}} \right)}{\cos \left( {{\varphi _{l,k}}} \right)}} { {\cos {\left( {{\phi _{k}}} \right)}}{\cos {\left( {{\varphi _{k}}} \right)}}} \\& {\hspace{-3cm} 0 {\le} {\varphi _{k}} {\le} {\Psi}} \\
0 & \hspace{-3cm}\rm{otherwise}, \\
\end{cases}
\vspace{-2mm}
\end{equation}
where $\tilde{r}$ is the reflection coefficient of the wall surface element $k$ with area $A_k$. The angles $\varphi_{l,k}$ and $\phi_{k}$ stand for the incident and irradiance angles in the surface element $k$, respectively. In contrast to the specular reflection, it is worth noticing that the distances are multiplied instead of added (see denominators in \eqref{eq:ChannelModelNLoSmirror}, \eqref{eq:ChannelModelNLoSRIS} and \eqref{eq:ChannelModelNLoSdiff}).

\begin{figure}[t]
     \centering
     \begin{subfigure}[b]{0.42\columnwidth}
         \centering
         \includegraphics[width=\textwidth]{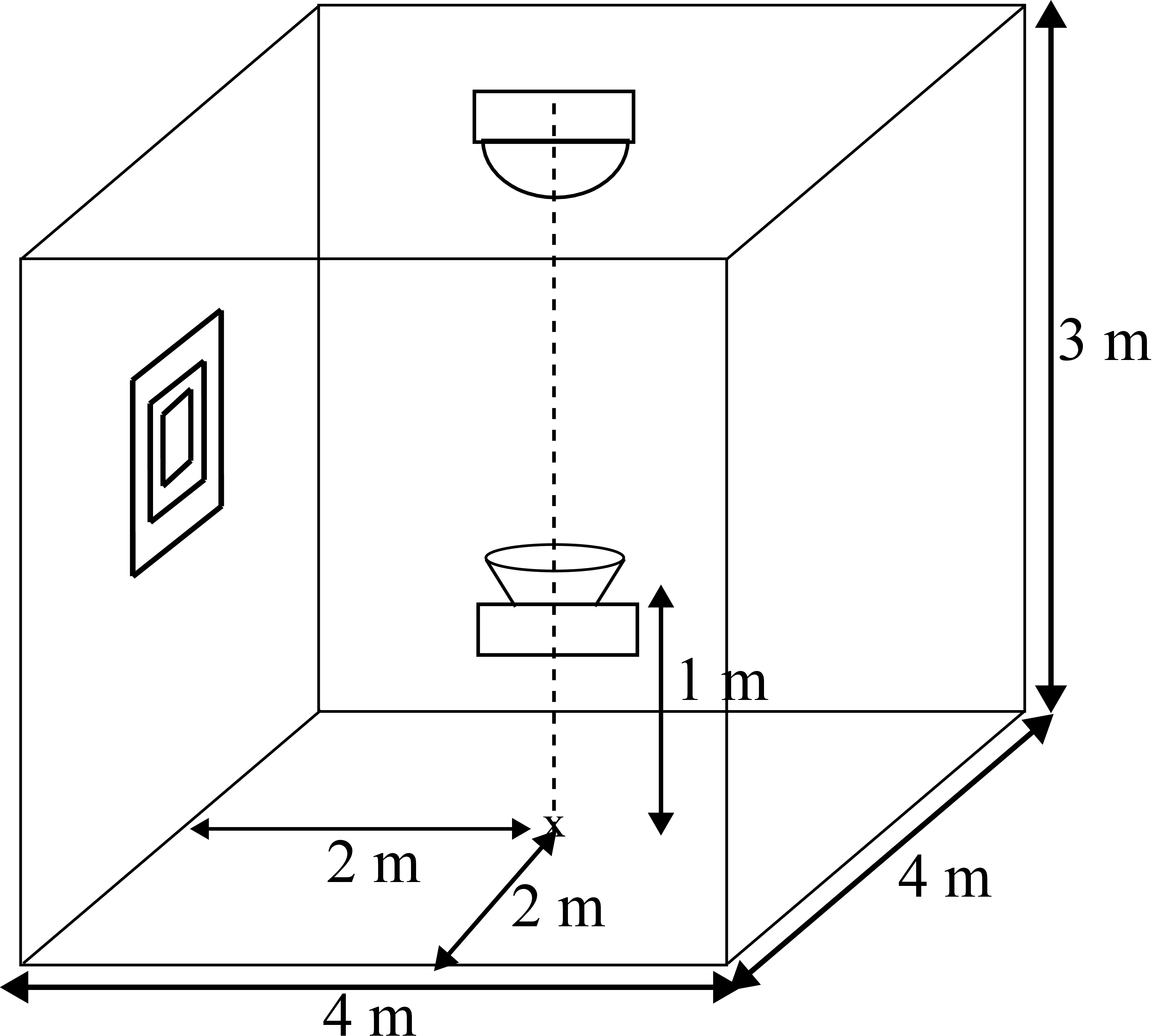}
         \caption{Scenario}
         \label{fig:ToyScenario}
     \end{subfigure}
     \hfill
          \begin{subfigure}[b]{0.56\columnwidth}
         \centering
         \includegraphics[width=\textwidth]{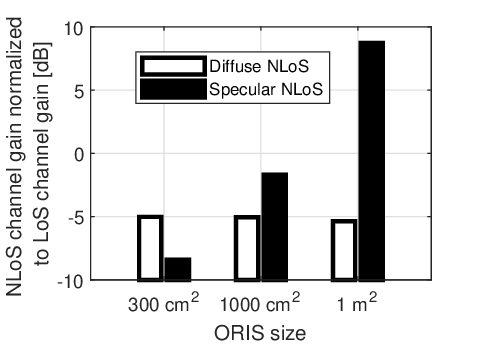}
         \caption{Channel gain at different ORIS sizes.}
         \label{fig:DiffvsSpecChannelGains}
     \end{subfigure}
        \caption{Comparison between diffuse and specular (ORIS) contributions with respect to LoS contribution.}        
        \label{fig:DiffvsSpec}
        \vspace{-4mm}
\end{figure}

Prior literature has demonstrated that, in an ORIS-aided VLC scenario, the diffuse contribution is negligible in comparison to the specular NLoS contribution and, consequently, only the specular NLoS reflections are considered~\cite{JointResourceMirrors}. However, this is not true for a small number of ORIS elements that create a small ORIS total size. Consider a scenario with the geometry represented in Fig.\,\ref{fig:ToyScenario}, where there is a single LED and a user, both located in the middle of the room. The parameters are $\phi_{1/2}=80^\circ$, $A_{\rm PD}=1\,$cm$^2$, $\tilde{r}=0.2$, $\hat{r}=0.99$ and the size of every ORIS element is 0.13x0.2 cm$^2$. Fig.\,\ref{fig:DiffvsSpecChannelGains} plots the channel gain contributions from the diffuse and specular reflections, both normalized to the LoS gain, when the ORIS have different total sizes. When the ORIS is small, the diffuse contribution is larger than the specular one, though the latter overcomes the diffuse one quickly as the total ORIS size increases. The specular reflection becomes even larger than the LoS contribution when the total ORIS size is large. In this paper, we aim to study the number and location of ORIS and mirror elements to obtain a minimum outage probability. This means that we are not setting up an ORIS structure by default that may be large enough so that the diffuse contributions may be disregarded. Therefore, in our study we must consider both diffuse and specular reflections. 

The second observation one can make from Fig.\,\ref{fig:DiffvsSpecChannelGains} is that the NLoS contributions are no longer negligible, as the NLoS gain can be significantly larger than the LoS gain in an ORIS-enhanced system. This is different from a non ORIS-aided VLC scenario, in which NLoS contributions have been typically disregarded when a LoS link exists~\cite{DLPerformanceAttocell}.

\vspace{-4mm}
\subsection{Overall VLC channel gain}

We model the reflector element placement with a binary variable $\beta_{l,k}\in\{0,1\},\, \forall l=0,1,\cdots,L{-}1,\, \forall k=0,1,\cdots,K{-}1$. The variable $\beta_{l,k}$ associates the LED $l$ with the reflector element $k$ to contribute to the user. It takes a value of 1 when the element $k$ is a specular reflector (ORIS or mirror), and a value of 0 when it is a diffuse reflector (wall). The NLoS channel gain from LED $l$ generated by the reflector element $k$ is thus formulated as 
\vspace{-1mm}
\begin{IEEEeqnarray}{rCl}
H_{l,k}^{{\rm{NLoS}}}\left( {{\beta _{l,k}}} \right) & {=} & H_{l,k}^{{\rm{wall}}}  {+} {\left( {H_{l,k}^{{\rm{spec}}} {-} H_{l,k}^{{\rm{wall}}}} \right)} {\cdot} {\beta _{l,k}}
\label{eq:ChannelModelNLoS}
\vspace{-1mm}
\end{IEEEeqnarray}
where the specular reflector can be either an ORIS or a mirror whose NLoS channel gain is defined as
\vspace{-2mm}
\begin{equation}
\label{eq:ChannelModelNLoSspec}
H_{l,k}^{{\rm{spec}}} =
\begin {cases}
H_{l,k}^{{\rm{ORIS}}} & {\text{if the specular reflector is an ORIS}}\\
H_{l,k}^{{\rm{mirror}}} & {\text{if the specular reflector is a mirror}}. \\
\end{cases}
\vspace{-1mm}
\end{equation}
Due to the intrinsic characteristic of specular reflections, every single element $k$ can forward the light power of at most a single LED, i.e.,
\vspace{-2mm}
\begin{IEEEeqnarray}{rCl}
\sum\limits_l {{{\beta _{l,k}}}  \le 1} ,\forall k.
\label{eq:BetaForSingleUserTx}
\vspace{-2mm}
\end{IEEEeqnarray}
Then, the overall VLC channel gain from LED $l$ to the user can be formulated as
\vspace{-2mm}
\begin{IEEEeqnarray}{rCl}
{H_{l}}\left( {{\beta _{l,k}}} \right) = {\rm I}_{l} \cdot H_{l}^{{\rm{LoS}}} + \sum\limits_k {{\rm I}_{l,k}\cdot H_{l,k}^{{\rm{NLoS}}}\left( {{\beta _{l,k}}} \right)},
\label{eq:ChannelModel}
\vspace{-2mm}
\end{IEEEeqnarray}
where ${\rm I}_{l}$ and ${\rm I}_{l,k}$ are Boolean variables that take a value of 0 when the LoS link from LED $l$ or NLoS link from LED $l$ passing through reflecting element $k$ are blocked, respectively, and 1 otherwise. 
Blockage may be generated by the user's own body (self-blocking), by other users, or by the presence of static objects such as furniture, pillars, etc. Therefore, the design of the reflector placement and the LED-reflector association denoted by $\beta_{l,k}$ are key to the system performance.

\vspace{-4mm}
\subsection{Mirror vs. ORIS NLoS contribution}
\label{subsec:MirrorVSRIS}
\begin{figure}[t]
     \centering
     \begin{subfigure}[b]{0.47\columnwidth}
         \centering
         \includegraphics[width=\textwidth]{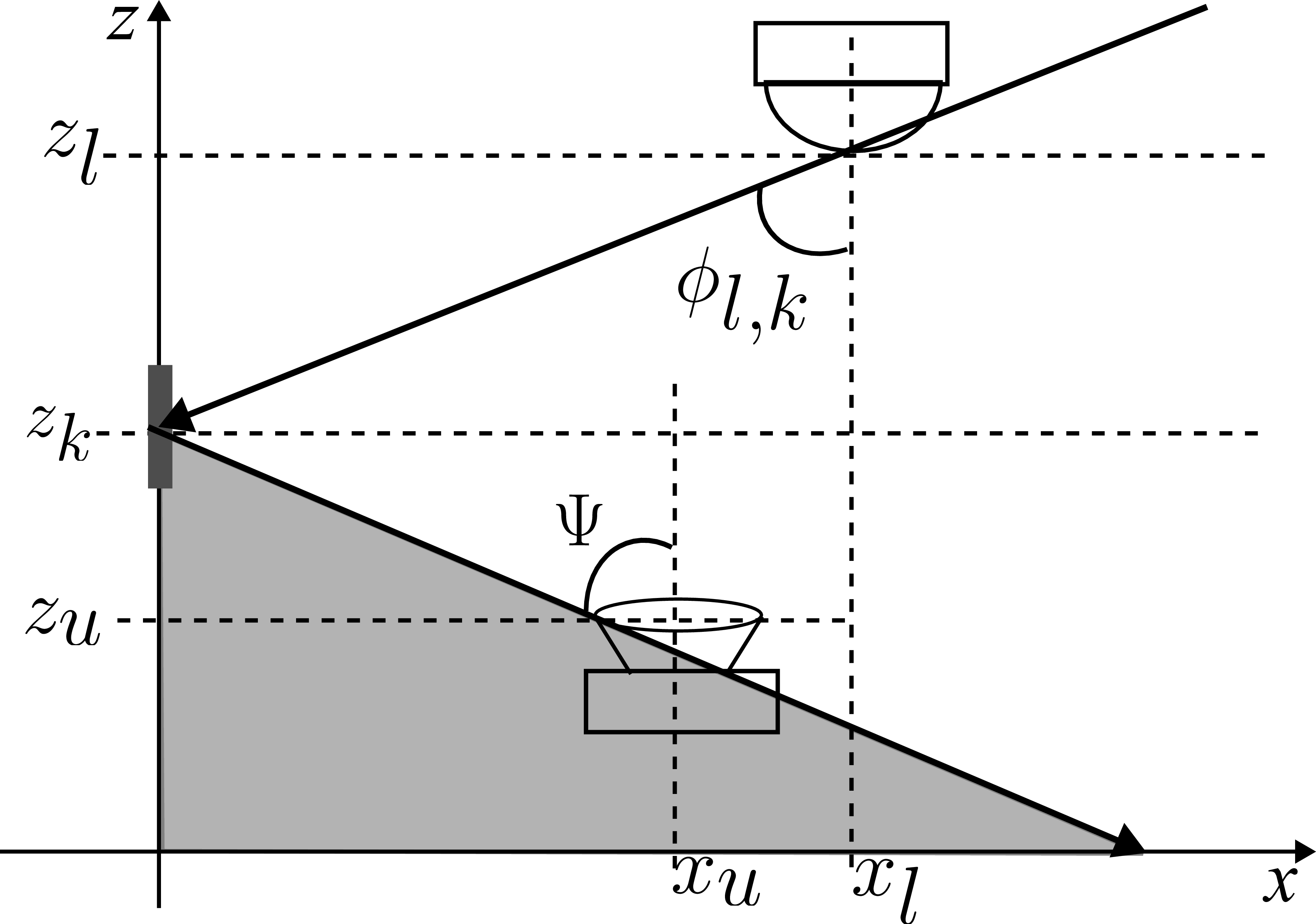}
         \caption{Mirror}
         \label{fig:GeometryMirror}
     \end{subfigure}
     \hfill
     \begin{subfigure}[b]{0.49\columnwidth}
         \centering
         \includegraphics[width=\textwidth]{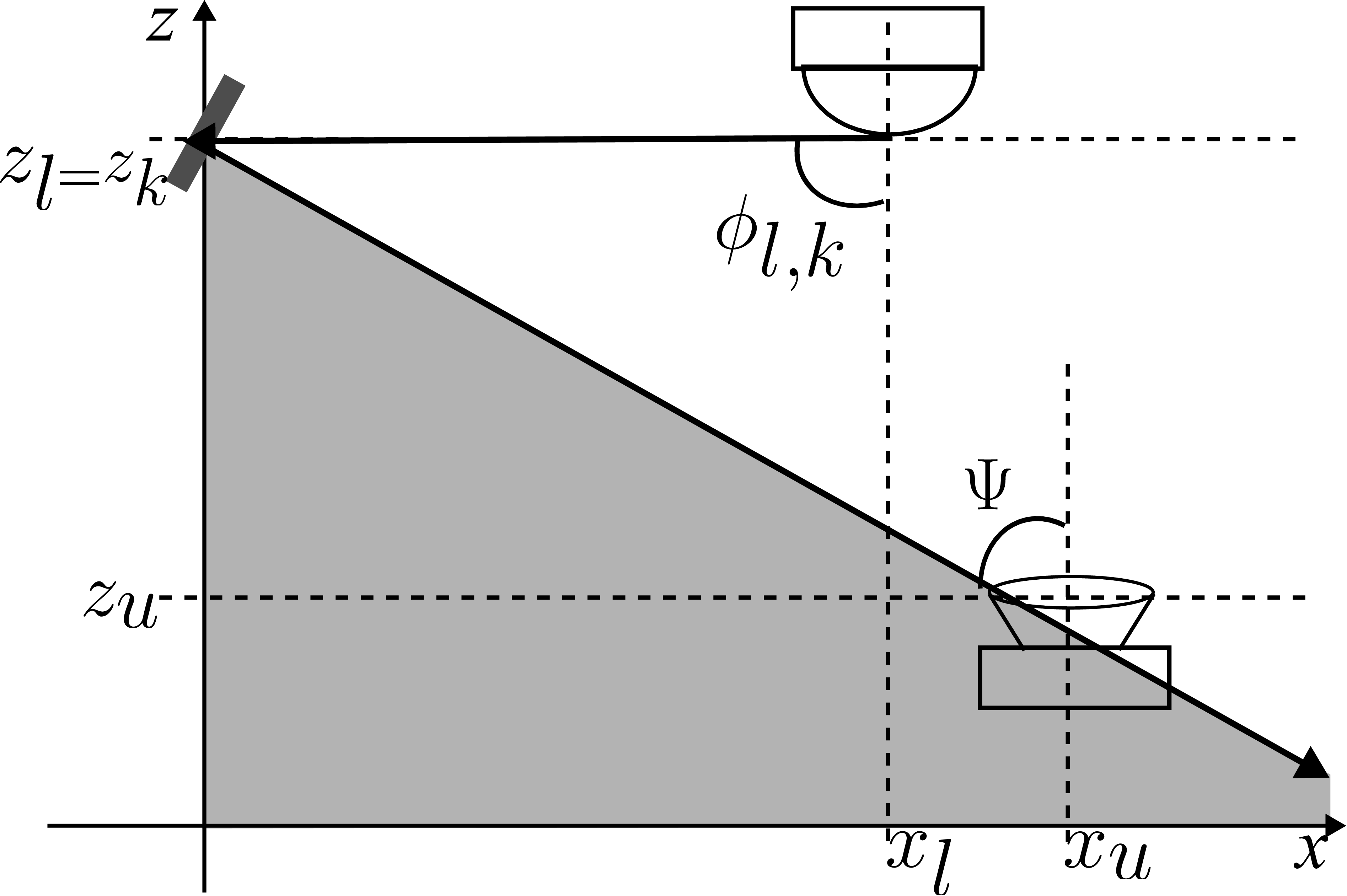}
         \caption{ORIS}
         \label{fig:GeometryRIS}
     \end{subfigure}
        \caption{Geometrical graphs of maximum distances reached by mirror and ORIS.}        
        \label{fig:Geometry}
        \vspace{-4mm}
\end{figure}

It is well known that an ORIS, due to its mobile orientation, potentially provides a greater communication performance than what a conventional mirror may achieve. This performance directly depends on the emission pattern of the LED, which is defined by its half-power semi-angle $\left(\phi_{1/2}\right)$, in addition to the PD FoV semi-angle $\left(\Psi\right)$ and the geometry of the scenario. For illustrative purposes, let us consider a 2D scenario with a single LED and a user's PD located at coordinates $\{x_l,z_l\}$ and $\{x_u,z_u\}$, respectively. All reflecting elements are installed in one wall and the location of the $k$-th reflecting element is determined by the coordinates $\{0,z_k\}$. The mirror and ORIS 2D geometric scenarios are \mbox{depicted in Fig.\,\ref{fig:Geometry}.}

In the case of a mirror, the maximum horizontal distance from the wall, $x_u$, that a user may occupy and still see the reflection is limited by the PD FoV semi-angle. According to \eqref{eq:ChannelModelNLoSmirror}, a non-zero NLoS mirror channel gain occurs when irradiance angle from LED $l$ to $k$ and incidence angle in the PD are the same. In the extreme case where the user is as far away from the wall as possible, we have $\Psi=\phi_{{l,k}}$. The shadowed area in Fig.\,\ref{fig:GeometryMirror} is where the user could be located to receive a NLoS contribution from a mirror. 
The area \mbox{is limited by} 
\vspace{-2mm}
\begin{IEEEeqnarray}{rCl}
z<-\tan\left({\frac{\pi}{2}-\Psi}\right)\cdot x + z_k.
\label{eq:LineMaxDistmirror}
\vspace{-2mm}
\end{IEEEeqnarray}
Substituting $z_k=z_l-x_l/\tan{\Psi}$ in \eqref{eq:LineMaxDistmirror}, we can formulate the maximum user distance as
\vspace{-2mm}
\begin{IEEEeqnarray}{rCl}
x_{u,{\rm max}}=\frac{z_u+\frac{x_l}{\tan{\Psi}}-z_l}{-\tan{\left(\frac{\pi}{2}-\Psi\right)}}.
\label{eq:MaxDistmirror}
\vspace{-2mm}
\end{IEEEeqnarray}
Note that the maximum horizontal distance with respect to the wall for the user so that it can receive contributions from the mirror depends on the PD height ($z_u$), the LED location $\{x_l,z_l\}$ and the PD FoV semi-angle ($\Psi$).

\begin{figure}[t]
     \centering
     \begin{subfigure}[b]{0.49\columnwidth}
         \centering
         \includegraphics[width=\textwidth]{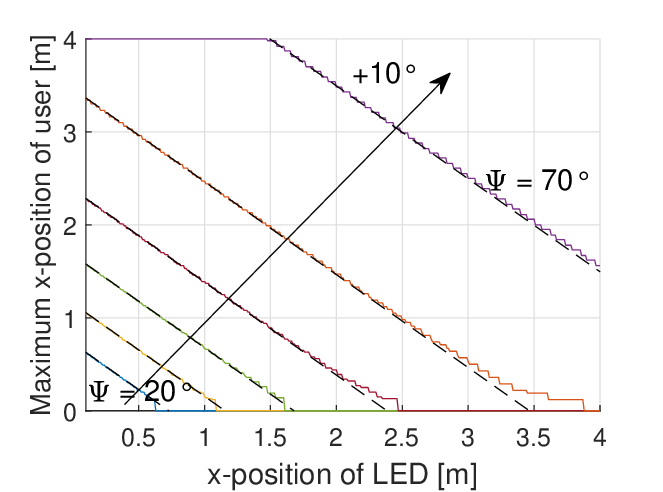}
         \caption{Mirror}
         \label{fig:MaxDist_FoVmirror}
     \end{subfigure}
     \begin{subfigure}[b]{0.49\columnwidth}
         \centering
         \includegraphics[width=\textwidth]{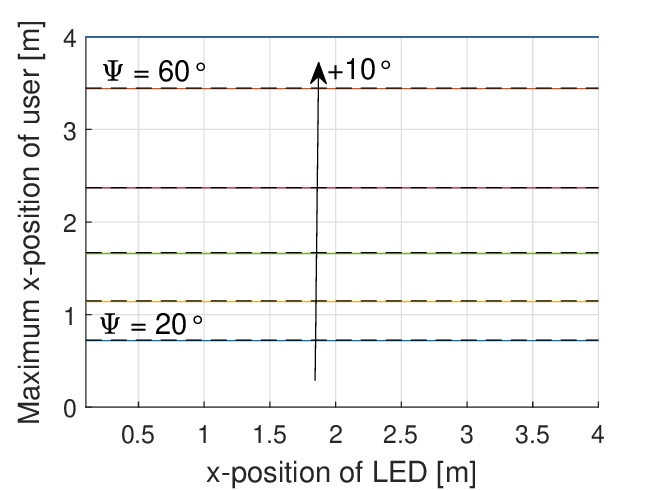}
         \caption{ORIS}
         \label{fig:MaxDist_FoVRIS}
     \end{subfigure}
        \caption{How $\Psi$ influences the maximum distance with respect to the wall where a user can be located to receive a NLoS contribution from a mirror or an ORIS. Parameters considered are $z_u$=1\,m and $z_l$=3\,m. Solid colored line: Simulation. Dashed black line: theoretical \eqref{eq:MaxDistmirror} and \eqref{eq:MaxDistRIS} for mirror and ORIS, respectively.}        
        \label{fig:MaxDist_FoV}
        \vspace{-5mm}
\end{figure}

Differently, in the case of an ORIS, since the mirror can be optimally oriented, the maximum horizontal distance, $x_u$, that a user may occupy and still receive some reflection is limited by the location of the highest ORIS, which is the location making $\varphi_{k}$ the smallest. In the most extreme case, we consider that ORIS elements may be placed as high as the ceiling. Then, the maximum $x$ distance from the wall where the user could receive a contribution from such ORIS is such that $\varphi_{k}=\Psi$, and it is formulated as
\vspace{-2mm}
\begin{IEEEeqnarray}{rCl}
x_{u,{\rm max}}=\left(z_l-z_u\right) \cdot \tan{\Psi}.
\label{eq:MaxDistRIS}
\end{IEEEeqnarray}
The shadowed area in Fig.\,\ref{fig:GeometryRIS} is the one where the user could be located to receive a NLoS contribution from an optimally oriented ORIS. Note that, unlike the mirror case, the maximum wall-user distance does not depend on the LED horizontal location ($x_l$) 
and thus, an ORIS can potentially cover a much larger area than a conventional mirror.

\subsubsection{Influence of FoV semi-angle $\left(\Psi\right)$} Let us consider a scenario where the user and LED are located at a height of $z_u=1$\,m and $z_l=3$\,m, respectively. The FoV semi-angle can be in the range of 20$^\circ$ to 70$^\circ$, and the LED can be located  ($x_l$) 0 to 4\,m from the wall containing the reflector. Figs.\,\ref{fig:MaxDist_FoVmirror} and \ref{fig:MaxDist_FoVRIS} plot the maximum x-position of a user ($x_{u,{\rm max}}$) for different LED positions $\left(x_l\right)$ and FoV semi-angles $\left(\Psi\right)$, when considering either a mirror or an ORIS as a reflecting surface, respectively. As expected from~\eqref{eq:MaxDistmirror}, 
the dependence of the maximum user distance on the LED's position limits the impact of using mirrors for a mirror-assisted VLC scenario. As an example, when the $\Psi=40^\circ$, the LED position must be at least 1.5\,m to support users that are located at a distance shorter than 1.5\,m from the wall, and thus a very small portion of the room area may be covered. We validate this analytical result from \eqref{eq:MaxDistmirror} and \eqref{eq:MaxDistRIS} for mirror and ORIS, respectively, (dashed lines in Figs.\,\ref{fig:MaxDist_FoVmirror} and \ref{fig:MaxDist_FoVRIS}) with simulations (solid lines). These results encourage the use of ORIS elements rather than conventional mirrors whenever possible. 

\subsubsection{Influence of half-power semi-angle $\left(\phi_{1/2}\right)$ in the ORIS case} 
\begin{figure}[t]
\centering
\includegraphics[width=\columnwidth]{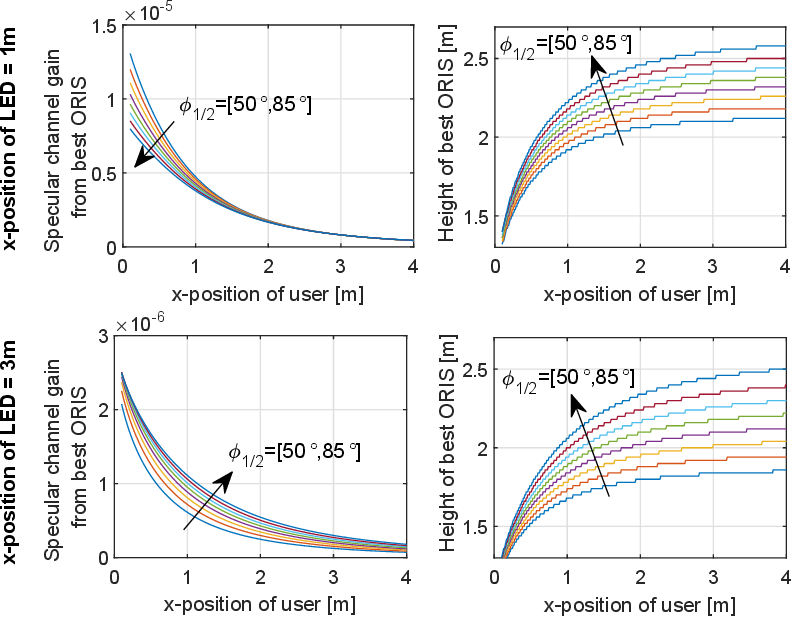}

\hspace{0.7cm} (a) \hspace{3.5 cm} (b) 
        \caption{How $\phi_{1/2}$ influences (a) the NLoS gain contribution and (b) the best ORIS placement for two LED placements of 1\,m and 3\,m from the wall containing the reflective element.}
\label{fig:NLoScontributionRIS_Phiangle}
\vspace{-4mm}
\end{figure}

Intuitively, the LED beamwidth determines the power impinging at each surface element, and then it has an effect on the NLoS contribution received by the user. Fig.\,\ref{fig:NLoScontributionRIS_Phiangle}a plots the $H_{l,k^*}^{{\rm{ORIS}}}$ gain, where $k^*$ is the reflecting element with the largest contribution and whose placement (height) is represented in Fig.\,\ref{fig:NLoScontributionRIS_Phiangle}b for $x_l=1$\,m and $x_l=3$\,m. Note that for short $x_l$ distances, the more directive the LED is, the better, whereas we can see the opposite effect for large $x_l$ values. 
However, differences obtained in channel gains are insignificant except when the LED is very separated from the wall, for which wider LED beams are preferable. This indicates that the selection of $\phi_{1/2}$ is not that important for the communication performance relying on specular NLoS, and that the $\phi_{1/2}$ configuration must be preferably selected according to illumination constraints. In Fig.\,\ref{fig:NLoScontributionRIS_Phiangle}b, the optimal height location varies depending on $\phi_{1/2}$ and the x-position of the user, and we observe that the more directive the LED is, the lower the ORIS must be located on the wall.

\vspace{-4mm}
\subsection{Figures of merit}
\vspace{-1mm}



Let us define $\mathbf{P}$ as the vector containing all $P_l,\, \forall l$ values, and $\boldsymbol{\beta}$ as the matrix containing all $\beta_{l,k}, \,\forall l,k$ values, respectively. Assuming that we transmit within the system bandwidth, the signal-to-noise power ratio (SNR) can be computed as 
\vspace{-4mm}
\begin{IEEEeqnarray}{l}
    {\gamma}({\mathbf{P}},{\boldsymbol{\beta}})  = 
    \frac{{{{\left( {{{\rho}\sum\limits_l\sum\limits_k P_l{H_{l}}\left( {{\beta _{l,k}}}\right)} } \right)}^2}}}{{{N_0}B}}, 
    \vspace{-1mm}
\label{eq:SNR}
\end{IEEEeqnarray}
where $\rho$ is the PD responsivity, $P_l$ is the optical power transmitted by LED $l$, $N_0$ is the power spectral density of the additive white Gaussian noise (AWGN) at the receiver mainly produced by shot and thermal noise~\cite{VLCLoSModel}, and $B$ is the communication bandwidth. 

The \emph{outage probability} is defined as the probability that a user has an SNR lower than the required SNR, called the SNR threshold ($\gamma_{{\rm th}}$), and can be written as
\vspace{-2mm}
\begin{IEEEeqnarray}{rCl}
P_{{\rm out}}({\mathbf{P}},{\boldsymbol{\beta}},\gamma_{{\rm th}}) & = &  \Pr\{\gamma({\mathbf{P}},{\boldsymbol{\beta}})<\gamma_{{\rm th}}\}  = \Gamma_{\gamma_{{\rm th}}},
\vspace{-2mm}
\label{eq:OutProb}
\end{IEEEeqnarray}
 where $\Gamma_{\gamma_{{\rm th}}}$ is the cumulative distribution function (CDF) of $\gamma({\mathbf{P}},{\boldsymbol{\beta}})$ evaluated at $\gamma_{{\rm th}}$. 

The \emph{optical energy efficiency} is defined as the data rate (measured in bit/s) that can be transmitted per unit of power (measured in W). This is equivalent to the number of bits that can be transmitted per unit of energy (J), expressed as bit/J and, invoking the tight lower bound of VLC system capacity~\cite{CapacityVLC}, it can be formulated as
\vspace{-2mm}
\begin{equation}
\label{eq:SpecEfficiency}
\eta_{{\rm opt}} =
\begin {cases}
\frac{\frac{B}{2}\log_2\left(1+\frac{\rm{exp(1)}}{2\pi}\cdot\gamma\right)}{\sum_lP_l} & {\text{if }} \gamma \geq \gamma_{\rm th}\\
0 & {\text{otherwise}}. \\
\end{cases}
\end{equation}


\section{Optimization Problem Formulation}\label{OptimizationProblem}
We aim to minimize the outage probability in a mirror- or ORIS-assisted VLC scenario while satisfying the illumination requirements in a room. For this, we propose the deployment of specular reflecting elements (mirrors or ORISs), which will allow us to understand to what extent they can enhance VLC performance.

\vspace{-4mm}
\subsection{Objective functions}
\label{subsec:ObjFunc}

The main goal is to minimize the outage probability defined as \eqref{eq:OutProb}. Let us refine the CDF of the SNR as $\Gamma_{\gamma_{{\rm th}}}(Q) = \Pr\{\gamma[{Q}, {\mathbf{P}(Q)},{\boldsymbol{\beta}(Q)}]<\gamma_{{\rm th}}\}$ considering a random variable $Q$ describing the user location and orientation. The random variable B$_{\gamma_{\rm th}}(Q)$ is an indicator function of $\gamma[Q,{\mathbf{P}(Q)},{\boldsymbol{\beta}(Q)}]$ defined as
\vspace{-2mm}
\begin{align}
\label{eq:BernoulliRandomVariable}
{\rm B}_{\gamma_{\rm th}}(Q) = 
\begin{cases}
0, & {\rm if}\,\gamma[Q,{\mathbf{P}(Q)},{\boldsymbol{\beta}(Q)}] < \gamma_{\rm{th}} \\
1, &  {\rm if}\,\gamma[Q,{\mathbf{P}(Q)},{\boldsymbol{\beta}(Q)}] \ge \gamma_{\rm{th}}
\end{cases},
    \vspace{-2mm}
\end{align}
where the LED distributed power $\mathbf{P}$ and the matrix of reflector element placements $\boldsymbol{\beta}$ depend only on the user location and orientation $Q$. 
We can now redefine the outage probability in \eqref{eq:OutProb} as  $\Pr\{{\rm B}_{\gamma_{\rm th}}(Q) =0\}$ = $\Pr\{\gamma\left[Q, \mathbf{P}(Q), \boldsymbol{\beta}(Q)\right]<\gamma_{\rm th}\}$. Instead of solving this mathematically, we opt to compute a sample average that can be solved by simulations. Then, 
\begin{IEEEeqnarray}{lr}
    \hspace{-3mm} \Pr\{\gamma{\left[Q, \mathbf{P}(Q), \boldsymbol{\beta}(Q)\right]}{\geq}\gamma_{\rm th}\}   & \IEEEnonumber\\ \hspace{-3mm}
    = \mathop{\mathbb{E}}\left[{\rm B}_{\gamma_{\rm th}}(Q)\right] {\approx} \frac{1}{N}\sum_{i=1}^N\Pr\{\gamma\left[{\rm q}_i, \mathbf{P}({\rm q}_i), \boldsymbol{\beta}({\rm q}_i)\right]\geq\gamma_{\rm th}\}, & 
\end{IEEEeqnarray}
where each ${\rm q}_i$ is a realization of $Q$. For a large $N$ value, this approximation approaches the mathematical solution. We optimize $\mathbf{P}({\rm q})$ and $\boldsymbol{\beta}({\rm q})$ for each q value, as we assume that the power allocation and mirror location can be adapted for each q value. Note that, in practice, varying the power allocation can be easily done, but varying the mirror location cannot. However, we aim to determine the mirror placement with the highest occurrence, which allows us to make such an assumption, and obtain an outage probability limit. That is, our results provide for the highest occurrence of locally optimal solutions as a guide to fixing the ORIS/mirrors placement.

Defining $b({\rm q})$ as a realization of the random variable ${\rm B}_{\gamma_{\rm th}}(Q)$  in \eqref{eq:BernoulliRandomVariable}, i.e.,
\begin{align}
\label{eq:outageevent}
b({\rm q}) =  {\rm B}_{\gamma_{\rm th}}(Q)\Big|_{{Q={\rm q}}} ,
\vspace{-2mm}
\end{align}
we may compute the set $\{[\mathbf{P}({\rm q}), \boldsymbol{\beta}({\rm q})]\}$ that maximizes $b({\rm q})$. Note that the set of functions $\{[\mathbf{P}({\rm q}), \boldsymbol{\beta}({\rm q})]\}$ that maximizes $b({\rm q})$ is typically infinite because it is binary valued. Considering this, as well as the management of available resources, we propose three objective functions ($\mathcal{O}$) that could be applied to our optimization problem:
\begin{itemize}
    \item $\mathcal{O}_1=b({\rm q})$, which focuses on the outage event of the user under study defined in \eqref{eq:outageevent}. 
    \item $\mathcal{O}_2=\sum_l\sum_k \beta_{l,k}({\rm q})$, which focuses on the number of mirror (or ORIS) elements deployed.
    \item $\mathcal{O}_3=\sum_lP_{l}({\rm q})$, which focuses on the amount of optical power allocated.
\end{itemize}

Ideally, we can formulate a multi-objective optimization problem that maximizes $\mathcal{O}_1$, while minimizing $\mathcal{O}_2$ and $\mathcal{O}_3$. However, when $\mathcal{O}_2$ increases, $\mathcal{O}_3$ may be decreased as lower power may be considered, and vice versa. Then, these are conflicting objective functions that cannot be optimized in conjunction, and it may lead, again, to an infinite number of Pareto-optimal solutions~\cite{Pareto}. To solve this issue we propose to define single-objective functions with some priorities, leading to a single optimal solution. Then, the outage event formulated in $\mathcal{O}_1$ is prioritized, and $\mathcal{O}_2$ and $\mathcal{O}_3$ are considered as regularization terms. Specifically, we propose the following optimization problem:
\begin{equation}
[\mathbf{\widehat{P({\rm q})}}, \widehat{\boldsymbol{\beta}({\rm q})}]  =  \argmax_{\mathbf{P({\rm q})}, \boldsymbol{\beta}({\rm q}),b({\rm q})} \mathcal{O}^*,
\label{eq:optimalPandBetaForGenericO}
\vspace{-2mm}
\end{equation}
where $\mathcal{O}^*$ can be either
\vspace{-2mm}\begin{IEEEeqnarray}{l}
\hspace{-4mm} \mathcal{O}_1^*{=}b({\rm q}){-}\epsilon{\cdot}\sum_l\sum_k \beta_{l,k}({\rm q}) \hspace{2mm}\text{or}\hspace{2mm} \mathcal{O}_2^*{=}b({\rm q}){-}\epsilon{\cdot}\sum_l P_l({\rm q}),
\,
\vspace{-2mm}
\end{IEEEeqnarray}
where $\epsilon$ is an infinitesimal value that forces the algorithm towards a unique optimal solution that minimizes the number of mirrors (or ORISs) employed in $\mathcal{O}_1^*$, or the amount of optical power allocated in $\mathcal{O}_2^*$, while prioritizing the outage event represented by variable $b$(q). For the remainder of the paper, we remove the functional dependence of $b$, $\mathbf{P}$ and $\boldsymbol{\beta}$ on q for notation simplicity as we consider the same objective functions for every user location and orientation.

An optimization problem can be solved with off-the-shelf software if constraints are linear, bilinear or quadratic, in both convex and non-convex optimization problems~\cite{Gurobi}. In the following, we analyze both communication and illumination constraints.

\subsection{Communication constraints}
\label{subsec:ConstrCommun}

We follow the big-$M$ approach to turn the conditional statement in \eqref{eq:BernoulliRandomVariable} into two linear inequalities~\cite{BigMApproach}, where $M$ is considered as an upper bound of $\gamma$ for every possible user location in the room:
\begin{IEEEeqnarray}{rCl}
    \hspace{-2mm}
{\gamma}({\mathbf{P}},{\boldsymbol{\beta}}) & \ge & {\gamma _{{\rm{th}}}} - M \cdot (1 - {b}),\label{eq:BigMapproach0}
\\
{\gamma}({\mathbf{P}},{\boldsymbol{\beta}}) & < & {\gamma _{{\rm{th}}}} + M \cdot {b}.
\label{eq:BigMapproach}
\end{IEEEeqnarray}
These inequalities define the $b$ variable, for our purpose, in such a way that both constraints are linear in $b$ when the user is in outage ($b=0$) and also when the user is not in outage ($b=1$). That is, when the optimal solution is that user is in outage, i.e., $b=0$, \eqref{eq:BigMapproach0} is satisfied because $M$ is a very large value to fulfill ${\gamma}({\mathbf{P}},{\boldsymbol{\beta}}) \ge {\gamma _{{\rm{th}}}} - M$, and \eqref{eq:BigMapproach} is also satisfied because ${\gamma}({\mathbf{P}},{\boldsymbol{\beta}}) < {\gamma _{{\rm{th}}}}$, which is the intrinsic definition of being in outage. Differently, when the optimal solution is that user is not in outage, i.e., $b=1$, \eqref{eq:BigMapproach0} is satisfied because
${\gamma}({\mathbf{P}},{\boldsymbol{\beta}}) \geq {\gamma _{{\rm{th}}}}$, which is the intrinsic definition of not being in outage, and \eqref{eq:BigMapproach} is also satisfied because $M$ is a very large value to fulfill ${\gamma}({\mathbf{P}},{\boldsymbol{\beta}}) < {\gamma _{{\rm{th}}}}+M$.

Note that ${\gamma}({\mathbf{P}},{\boldsymbol{\beta}})$ is neither linear, nor bilinear, nor quadratic with respect to the variables $P_l$ and $\beta_{l,k}$ (see \eqref{eq:SNR}), since the summation in the numerator is squared, leading to cubic terms. To simplify its formulation, we create a new variable that denotes the multiplication between the optical power transmitted by LED $l$, i.e. $P_l$, and the \mbox{binary variable $\beta_{l,k}$:} 
\begin{IEEEeqnarray}{rCl}
{\varrho _{l,k}} & = & {P_l} \cdot {\beta_{l,k}},
\label{eq:varrho}
\end{IEEEeqnarray}
which leads to the matrix $\boldsymbol{\varrho}$ containing all ${\varrho _{l,k}}\,\forall l,k$ values. Thus, the SNR in \eqref{eq:SNR} can be re-formulated as in \eqref{eq:SNReq} at the top of the next page,
where the formulation between curly brackets can be re-written as the sum of terms listed in Table\,\ref{tab:DegreePolynomicSNR}, whose degrees with respect to the variables $P_l$ and $\varrho_{l,k}$ are also included.
As we can see, these terms and therefore, all communication constraints, are no longer cubic, and they can be formulated and solved with off-the-shelf software.

\begin{figure*}[t]
\begin{IEEEeqnarray}{l}
\label{eq:SNReq}
{\gamma}\left( {\mathbf{P}},{\boldsymbol{\varrho}} \right) =   
  \frac{{\rho^2}}{{{N_0}B}} {\cdot} \Bigg\{ \hspace{-1mm}\sum\limits_l \hspace{-1mm} {{{\left[ {{I_{l}} {\cdot} P_l{\cdot} H_{l}^{{\rm{LoS}}} {+} \hspace{-1mm} \sum\limits_k {{I_{l,k}} {\cdot} \left(P_l{\cdot}H_{l,k}^{\rm wall}+\varrho_{l,k}{\cdot}\left(H_{l,k}^{\rm spec}-H_{l,k}^{\rm wall} \right)\right)} } \right]}^2}}  \IEEEyesnumber\\  + 2 {\cdot} \hspace{-1mm}\sum\limits_{l'=1}^{L-1} \hspace{-0.5mm}\sum\limits_{l=0}^{l'-1} \hspace{-1mm} {\Bigg[ \hspace{-0.5mm} {{I_{l}}  {\cdot} \hspace{-0.5mm} \underbrace{P_l{\cdot} H_{l}^{{\rm{LoS}}}}_\text{$R_{l}^{{\rm{LoS}}}(P_l)$} \hspace{-0.5mm} {+} \hspace{-1mm} \sum\limits_k \hspace{-1mm}{{I_{l,k}} {\cdot} \hspace{-1mm} \underbrace{\left(\hspace{-0.5mm}P_l{\cdot}H_{l,k}^{\rm wall}{+}\varrho_{l,k}{\cdot} \hspace{-1mm} \left(\hspace{-0.5mm}H_{l,k}^{\rm spec}\hspace{-0.5mm}{-}H_{l,k}^{\rm wall} \right)\hspace{-0.5mm}\right)}_\text{$R_{l,k}^{{\rm{NLoS}}}(P_l,\varrho_{l,k})$} } } \Bigg]} \hspace{-0.5mm} {\cdot} \hspace{-0.5mm} {\Bigg[ {{I_{l'}} {\cdot} \hspace{-0.5mm} \underbrace{P_{l'}{\cdot} H_{l'}^{{\rm{LoS}}}}_\text{$R_{l'}^{{\rm{LoS}}}(P_l')$} \hspace{-0.5mm}{+} \hspace{-1mm} \sum\limits_k \hspace{-1mm}{{I_{l',k}} {\cdot} \hspace{-1mm} \underbrace{\left(\hspace{-0.5mm}P_{l'}{\cdot}H_{l',k}^{\rm wall}{+}\varrho_{l',k}{\cdot} \hspace{-1mm} \left(\hspace{-0.5mm} H_{l',k}^{\rm spec}\hspace{-0.5mm}{-}H_{l',k}^{\rm wall} \right)\hspace{-0.5mm}\right)}_\text{$R_{l',k}^{{\rm{NLoS}}}(P_{l'},\varrho_{l',k})$}} } \hspace{-0.5mm} \Bigg]}  \hspace{-0.5mm} \Bigg\} \IEEEnonumber
    \vspace{-2mm}
\end{IEEEeqnarray}
\rule{\textwidth}{2pt}
\vspace{-8mm}
\end{figure*}

\setlength{\tabcolsep}{2pt}
\begin{table}[t]
\vspace{0.05in}
    \centering
     \caption{Terms of $\gamma$ in~\eqref{eq:SNReq}, and their degrees (Qu: quadratic, Bi: bilinear) with respect to $P_l$ and $\varrho_{l,k}$. }
    \begin{tabular}{|c|c|}
    \hline
    \rowcolor{light_grey} 
     \textbf{Term} & \textbf{Degree} \\
    \hline
      $\sum\limits_l {{I_{l}}}  \cdot {\left( {R_{l}^{{\rm{LoS}}}\left( {{P_l}} \right)} \right)^2}$ & Qu\\
        \hline
      $\sum\limits_l {{{\sum\limits_k {{I_{l,k}} \cdot \left( {R_{l,k}^{{\rm{NLoS}}}\left( {{P_l},{\varrho _{l,k}}} \right)} \right)} }^2}}$ & Qu, Bi\\
         \hline
      $2\sum\limits_l \hspace{-0.5mm} {\sum\limits_{k'=1}^{K{-}1} \hspace{-0.5mm}\sum\limits_{k=0}^{k'{-}1} \hspace{-1mm} {{I_{l,k}} \hspace{-0.5mm} {\cdot} \hspace{-0.5mm} {I_{l,k'}} \hspace{-0.5mm} {\cdot} \hspace{-0.5mm}  {R_{l,k}^{{\rm{NLoS}}}\hspace{-1mm}\left( {{P_l},\hspace{-0.5mm}{\varrho _{l,k}}} \right)}   {\cdot}   {R_{l,k'}^{{\rm{NLoS}}}\hspace{-1mm}\left( {{P_l},\hspace{-0.5mm}{\varrho _{l,k'}}} \right)} } }$ & Qu, Bi\\
         \hline
      $2 \sum\limits_l {{I_{l}}}  {\cdot} R_{l}^{{\rm{LoS}}}\left( {{P_l}} \right) {\cdot} \sum\limits_k {{I_{l,k}} {\cdot} R_{l,k}^{{\rm{NLoS}}}\left( {{P_l},{\varrho _{l,k}}} \right)}$ & Qu, Bi \\
         \hline
      $2\sum\limits_{l'=1}^{L-1} \hspace{-0.5mm}\sum\limits_{l=0}^{l'-1} {{I_{l}} \cdot {I_{l'}} \cdot R_{l}^{{\rm{LoS}}}\left( {{P_l}} \right) \cdot R_{l'}^{{\rm{LoS}}}\left( {{P_{l'}}} \right)}$ & Bi \\
         \hline
      $2 \sum\limits_{l'=1}^{L-1} \hspace{-0.5mm}\sum\limits_{l=0}^{l'-1} {{I_{l}} {\cdot} R_{l}^{{\rm{LoS}}}\left( {{P_l}} \right) {\cdot} \sum\limits_k {{I_{l',k}} {\cdot} R_{l',k}^{{\rm{NLoS}}}\left( {{P_{l'}},{\varrho _{l',k}}} \right)} }$ & Bi \\
         \hline
      $ 2 \sum\limits_{l'=1}^{L-1} \hspace{-0.5mm}\sum\limits_{l=0}^{l'-1} {{I_{l'}} {\cdot} R_{l'}^{{\rm{LoS}}}\left( {{P_{l'}}} \right) {\cdot} \sum\limits_k {{I_{l,k}} {\cdot} R_{l,k}^{{\rm{NLoS}}}\left( {{P_l},{\varrho _{l,k}}} \right)} }$ & Bi\\
         \hline      
         $2 \hspace{-1mm} \sum\limits_{l'=1}^{L-1} \hspace{-0.5mm}\sum\limits_{l=0}^{l'-1}  \hspace{-1mm}\Bigg(\hspace{-1mm}\sum\limits_k \hspace{-1mm} {{I_{l,k}} {\cdot} R_{l,k}^{{\rm{NLoS}}}\hspace{-0.5mm}\left( {{P_l},{\varrho _{l,k}}} \right)}\hspace{-1mm}\Bigg) \hspace{-0.5mm}{\cdot} \hspace{-0.5mm}\Bigg(\hspace{-1mm}\sum\limits_k \hspace{-0.5mm} {{I_{l',k}} {\cdot} R_{l',k}^{{\rm{NLoS}}}\hspace{-0.5mm}\left( {{P_{l'}},{\varrho _{l',k}}} \right)} \hspace{-1mm}\Bigg)$ & Bi\\
    \hline
        \end{tabular}
    \label{tab:DegreePolynomicSNR}
    \vspace{-4mm}
\end{table}
\setlength{\tabcolsep}{6pt}

\vspace{-4mm}
\subsection{Illumination constraints}
\label{subsec:ConstrIllum}
The design of a VLC network must pay special attention to not infringe on the primary functionality of a lighting system. First, the \emph{average illuminance} $E_{{\rm avg}}$ in the whole area must be larger than a threshold $E_{{\rm th}}$, i.e., $E_{{\rm avg}} \geq E_{{\rm th}}$, which is determined by the type of activity carried out in the space. For example, a  library requires a larger average illuminance than a residential space~\cite{LightingStandard}. The average illuminance in a room is formulated as
\vspace{-2mm}
\begin{IEEEeqnarray}{rCl}
E_{\rm avg} & = &  \frac{1}{N}\cdot \sum_n E_{\rm v}(n),
\label{eq:Eavg}
\vspace{-2mm}
\end{IEEEeqnarray}
where $n=\{0,...,N-1\}$ are all the possible sensing points that cover the whole room area, and $E_{\rm v}(n)$ is the illuminance at point $n$ defined as
\vspace{-2mm}
\begin{IEEEeqnarray}{rCl} 
E_{\rm v}(n) & = &  \frac{K_{\rm e/v}}{A_{\rm PD}}\cdot \sum_l P_lH_{l}^{\rm LoS}(n),
\label{eq:En}
\vspace{-2mm}
\end{IEEEeqnarray}
where $K_{\rm e/v}$ (measured in lm/W) is the luminous efficacy of the white light that the LED generates, and $H_{l}^{\rm LoS}(n)$ is the LoS channel gain from LED $l$ to the room point $n$, computed as~\eqref{eq:ChannelModelLoS}. Note that the illuminance is computed only with the LoS contribution, since illumination must be guaranteed regardless of reflections; it cannot depend on the wall material and color, which can change as the space is decorated after the lighting has been installed.

The second illumination constraint to fulfill is that the illuminance at every point $n$ must be lower than or equal to a maximum value $E_{{\rm max}}$, i.e., $E_{{\rm v}}(n)\leq E_{{\rm max}},\,\forall n$. This constraint is required for eye safety purposes. 

The third and last illumination constraint to satisfy is that the \emph{lighting uniformity} $U$ must be larger than a minimum $U_{{\rm min}}$, i.e. $U\geq U_{{\rm min}}$, where the lighting uniformity is defined as
\vspace{-2mm}
\begin{IEEEeqnarray}{rCl} 
U & = &  \frac{\min_n E_{\rm v}(n)}{E_{\rm avg}}.
\label{eq:Uniformity}
\vspace{-2mm}
\end{IEEEeqnarray}
Note that $U$ is not a linear function with respect to the variable $P_l$. To linearize $U$ we formulate a new variable $E_{\rm min}=\min_n E_{\rm v}(n)$, which leads to the following two new constraints
\vspace{-2mm}
\begin{equation}
E_{\rm min}  \geq  U_{\rm min}\cdot E_{\rm avg} \qquad \text{and} \qquad
E_{\rm min}  \leq  E_{\rm v}(n),\,\forall n.
\label{eq:UniformityIneq}
\end{equation}

\vspace{-4mm}
\subsection{Solvable optimization problem}

Note that the objective functions in Section\,\ref{subsec:ObjFunc} are linear and, together with the re-arrangements introduced in Section\,\ref{subsec:ConstrCommun} and Section\,\ref{subsec:ConstrIllum}, our optimization problem can be formulated as a mixed-integer programming (MIP) problem with linear, bilinear and quadratic constraints. It now complies with all features to be solved by a non-convex solver such as Gurobi and CVX interface~\cite{Gurobi, cvx}. We formulate our \textsc{JointMinOut} optimization problem as
\vspace{-2mm}
\begin{IEEEeqnarray}{l}
\label{MINOUT}
\textsc{JointMinOut:} \mathop {\mathop {\max }\limits_{{\boldsymbol{\beta}},{b},{\mathbf{P}},{E_{\rm min }}} \mathcal{O}_1^* \text{ (or }\mathcal{O}_2^*) }\limits\\
\begin{array}{l}
{\rm{subject\,to}}\\
\rotatebox[origin =c]{90}{\parbox{3.5 em}{\footnotesize \centering Resource\\ constraints}}
\begin{cases}
{\rm{C1:  }}\,\sum\limits_l \sum\limits_k  {{\beta _{l,k}} \le {N_{{\rm{max}}}}} \\
{\rm{C2:  }}\,\sum\limits_l { {{\beta _{l,k}}}  \le 1} ,\forall k 
\end{cases}\\
\rotatebox[origin =c]{90}{\parbox{3.5 em}{\footnotesize \centering Commun.\\ constraints}}
\begin{cases}
{\rm{C3:  }}\,{\gamma}({\mathbf{P}},{\boldsymbol{\varrho}}) \ge {\gamma _{{\rm{th}}}} - M \cdot (1 - {b}),{\rm{  }}\\
{\rm{C4:  }}\,{\gamma}({\mathbf{P}},{\boldsymbol{\varrho}}) < {\gamma _{{\rm{th}}}} + M \cdot {b},{\rm{  }}
\end{cases}\\
\rotatebox[origin =c]{90}{\parbox{6 em}{\footnotesize \centering Illumination\\ constraints}}
\begin{cases}
{\rm{C5:  }}\,\frac{{{K_{{\rm{e/v}}}}}}{{N{A_{{\rm{PD}}}}}} \cdot \sum\limits_l {\sum\limits_n {P_lH_{l}^{\rm LoS}(n) \ge {E_{{\rm{th}}}}} } \\
{\rm{C6:  }}\,{E_{\min }} \ge {{{U}}_{\rm{min}}} \cdot \frac{{{K_{{\rm{e/v}}}}}}{{N{A_{{\rm{PD}}}}}} \cdot \sum\limits_l {\sum\limits_n {P_lH_{l}^{\rm LoS}(n)} } \\
{\rm{C7:  }} \,{E_{\min }} \le \frac{{{K_{{\rm{e/v}}}}}}{{{A_{{\rm{PD}}}}}} \cdot \sum\limits_l {P_lH_{l}^{\rm LoS}(n)}, \forall n\\
{\rm{C8:  }} \, \frac{{{K_{{\rm{e/v}}}}}}{{{A_{{\rm{PD}}}}}} \cdot \sum\limits_l {P_lH_{l}^{\rm LoS}(n)} \le {E_{\max }} , \forall n
\end{cases}\\
\rotatebox[origin =c]{90}{\parbox{6 em}{\footnotesize \centering Definition of\\ variables}}
\begin{cases}
{\rm{C9:  }}\,{\varrho _{l,k}} = {P_l} \cdot {\beta _{l,k}},{\rm{  }}\forall l,k\\
{\rm{C10:  }}\,{P_l} \ge 0,{\rm{  }}\forall l\\
{\rm{C11:  }}\,{\beta _{l,k}} \in \{ 0,1\} ,{\rm{  }}\forall l,k\\
{\rm{C12:  }}\,{b} \in \{ 0,1\}\\
{\rm{C13:  }}\,{E_{\min }} \ge 0\IEEEnonumber
\end{cases}\\
\end{array} 
\vspace{-2mm}
\end{IEEEeqnarray}

We maximize the objective function subject to the following constraints: C1-C2 refer to the maximum possible number of mirrors (or ORISs) given by $N_{\rm max}$, and to the LED-mirror association defined in \eqref{eq:BetaForSingleUserTx};  
 C3-C4 invoke the Big-$M$ approach to define the outage variable $b$;
 C5 defines the average illuminance constrained at a minimum value of $E_{\rm th}$; 
 C6 defines the illuminance uniformity constrained at a minimum value of $U_{{\rm min}}$; 
 C7 refers to the minimum illuminance $E_{{\rm min}}$ along the whole room required to define illuminance uniformity in C6; 
 C8 defines the maximum illuminance along the whole room; 
 and C9-13 include definitions for all the variables used in the optimization problem.

If we assume the variable $P_l$ is known and fixed for all LEDs, C5-C10 and C13 are not needed, and the problem becomes a 0-1 integer linear program. This is one of Karp's 21 NP-complete problems and requires exponential time with respect to the input size to solve~\cite{Karp1972}. 
We can solve this for small enough problems and with enough computational resources, but the problem becomes intractable when considering realistic scenarios. The input size equals $K\cdot L$, which is very large as the value of $K$ must be in the hundreds so that the walls are partitioned into sufficiently small elements to get an accurate NLoS channel representation. Including an unknown power allocation $P_l,\,\forall l$ as a new variable makes the problem even more difficult to solve, requiring heuristic approaches.

\vspace{-2mm}
\section{Proposed heuristic algorithms}\label{AlgorithmsProposed}

We propose two heuristic alternating iterative algorithms to solve the \textsc{JointMinOut} problem, and we compare their performance to the benchmark that can be formulated as a non-iterative optimization algorithm.

\vspace{-4mm}
\subsection{Proposed alternating optimization (AO) algorithms}

We have two resources that must be studied: number of mirrors (or ORIS) and their placement, which is defined by the parameter $\boldsymbol{\beta}$, and power allocated to each LED $l$, defined by the parameter $\mathbf{P}$. To solve the \textsc{JointMinOut} optimization problem detailed in \eqref{MINOUT}, we propose an alternating optimization (AO) algorithm that divides the problem into two subroutines to find the global optimization variables $\mathbf{P}$ and $\boldsymbol{\beta}$ iteratively, subject to fixing one of them each time, until the problem converges. 

\begin{figure}[t]
\centering
\includegraphics[width=\columnwidth]{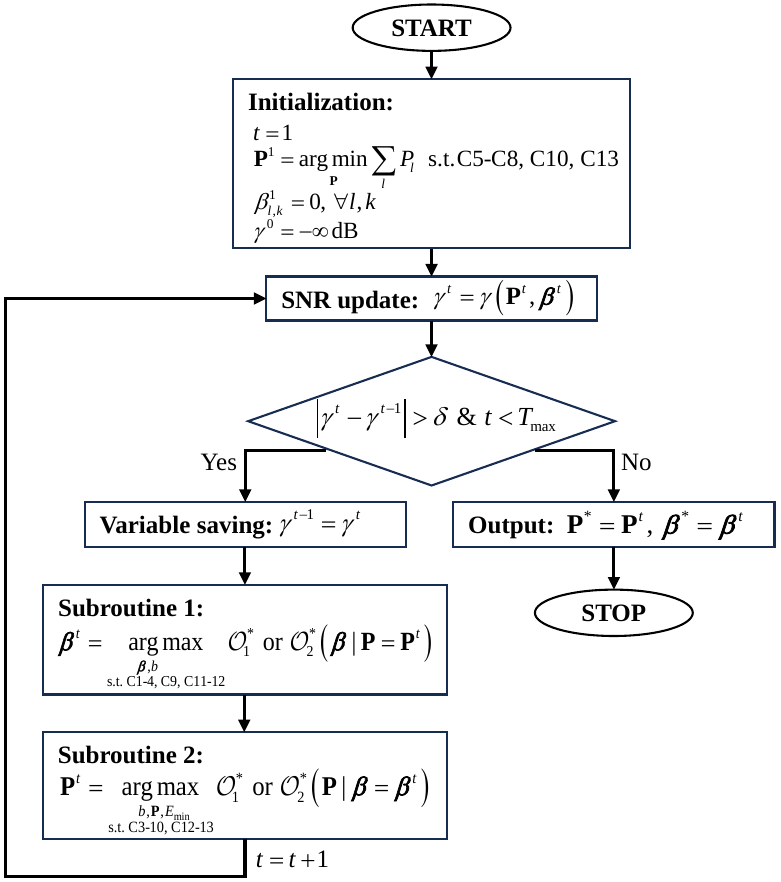}
        \caption{Proposed generalized alternating optimization algorithm.}
\label{fig:AOalgorithmFlowGraph}
\vspace{-5mm}
\end{figure}

A flow diagram of the general AO algorithm proposed is included in Fig.\,\ref{fig:AOalgorithmFlowGraph}. The algorithm starts with the initialization of variables for iteration $t=1$: the output power for each LED $l$ contained in $\mathbf{P}^1$, is initially set as the minimum power required to comply with the illumination constraints defined in \eqref{MINOUT}; and the algorithm is initialized with no mirrors deployed, i.e. $\beta_{l,k}^1=0, \forall l,k$. The two variables under study at iteration $t$ are denoted as $\mathbf{P}^t$ and $\boldsymbol{\beta}^t$. Then, at each iteration, the SNR value $\gamma^t$ is updated to be compared with the value of the SNR in the previous iteration, $\gamma^{t-1}$. The variables $\boldsymbol{\beta}^t$ and $\mathbf{P}^t$ are iteratively updated as follows:
\begin{itemize}
    \item Subroutine 1 computes $\boldsymbol{\beta}^t$ that maximizes the function $\mathcal{O}_1^*$ (or $\mathcal{O}_2^*$) for a given $\mathbf{P}=\mathbf{P}^t$. Only the constraints depending on $\boldsymbol{\beta}^t$ are considered, which correspond to resource and communication constraints (C1-4), and some definitions (C9, C11-12). 
    \item Subroutine 2 computes $\mathbf{P}^t$ that maximizes the function $\mathcal{O}_1^*$ (or $\mathcal{O}_2^*$) for a given $\boldsymbol{\beta}=\boldsymbol{\beta}^t$ obtained in subroutine 1. Only constraints depending on $\mathbf{P}$ are considered, which correspond to C3-10 and C12-13.
\end{itemize}
The algorithm leaves the loop when the problem converges, i.e. the SNR changes less than an infinitesimal $\delta$ value with respect to the previous iteration, or when the number of iterations $t$, i.e., execution time, exceeds $T_{\rm max}$. 
Depending on the resource to conserve, either the number of mirrors (or ORISs) or the amount of power allocated, we perform $\mathcal{O}_1^*$ or $\mathcal{O}_2^*$, respectively. Both focus on maximizing $b$, i.e., prioritizing solutions for which $b=1$ and the user is not in outage. 

Subroutine 1 is an integer linear programming problem where the unknown variables are binary. Although we have simplified the problem with respect to \textsc{JointMinOut}, this is still NP-complete. 
Thus, we propose two heuristic algorithms that replace subroutine 1, one for each resource to conserve (number of mirrors or optical power):

\subsubsection{Minimum Mirrors (MM) approach}

This approach minimizes the number of mirrors (or ORIS) employed, then allocating as much power as the constraints allow and exploiting mirrors (or ORISs) only as a last resort. The pseudo-code for the heuristic MM approach to solve subroutine 1 is described in Algorithm\,\ref{alg:Subroutine1MM}. The variables required to perform the algorithm are $N_{{\rm max}},\,I_{l,k},\, H_{l,k}^{\rm spec}$ and $P_l^{t},\, \forall l,k$. 
The algorithm computes the optimal $\{l,k\}$ that satisfies $\gamma(\mathbf{P}^{t},\boldsymbol{\beta}^t){>}\gamma_{{\rm th}}$ while minimizing the number of mirrors (or ORISs) that contribute to increasing the SNR. 

\begin{algorithm}[t]
\caption{Pseudo-code for subroutine 1 for the MM approach}\label{alg:Subroutine1MM}
\KwData{$N_{{\rm max}},\,I_{l,k},\, H_{l,k}^{\rm spec},\,P_l^{t},\, \forall l,k$}
\KwResult{$\boldsymbol{\beta}^t$}
      \If{$I_{l,k}H_{l,k}^{\rm spec}P_l^{t}\neq \mathbf{0}$}
    {
      $N=0$\\
      \While{$\gamma({\mathbf{P}^t},{\boldsymbol{\beta}^t}){<}\gamma_{{\rm th}}$ {{\rm and}} $N{\leq} N_{{\rm max}}$}{
$\{l^*,k^*\}=\hspace{-10mm}\argmax\limits_{\substack{l\in L, k\in K'\\\hspace{10mm}{{\rm s.t.}} K'=\left\{k\in K \big||K'|=N\right\} \\ \hspace{4mm}\sum_l \beta_{l,k}=1,\forall k}} \hspace{-10mm} \sum\limits_{l}\sum\limits_{k}I_{l,k}H_{l,k}^{\rm spec}P_l^{t}$\\
$\beta_{l,k}^t=1,\forall \{l,k\}\in \{l^*,k^*\}$\\
$N=N+1$}}
\end{algorithm}

\subsubsection{Minimum Power (MP) approach} 
As a commitment to the environment, this approach minimizes the power transmitted by the LEDs while exploiting mirrors (or ORIS) as much as possible. 
The pseudo-code describing the heuristic MP approach replacing subroutine 1 is described in Algorithm\,\ref{alg:Subroutine1MP}. The MP approach selects the $N_{{\rm max}}$ best $\{l,k\}$-pairs that maximize the total received reflection so that the transmitted power can be minimized. 


In both cases, the mirror (or ORIS) selection $\boldsymbol{\beta}^t$ obtained in subroutine 1 is used in subroutine 2, which can be solved by simple linear programming optimization.

\begin{algorithm}[t]
\caption{Pseudo-code for subroutine 1 for the \mbox{MP approach}}\label{alg:Subroutine1MP}
\KwData{$N_{{\rm max}},\,I_{l,k},\, H_{l,k}^{\rm spec},\,P_l^{t},\, \forall l,k$}
\KwResult{$\boldsymbol{\beta}^t$}
      \If{$I_{l,k}H_{l,k}^{\rm spec}P_l^{t}\neq \mathbf{0}\, \forall l,k$}
{$\{l^*,k^*\}=\hspace{-10mm}\argmax\limits_{\substack{l\in L, k\in K'\\\hspace{12mm}{{\rm s.t.}} K'=\left\{k\in K \big||K'|=N_{\rm max}\right\} \\ \hspace{3mm}\sum_l \beta_{l,k}=1,\forall k}} \hspace{-10mm} \sum\limits_{l}\sum\limits_{k}I_{l,k}H_{l,k}^{\rm spec}P_l^{t}$\\
$\beta_{l,k}^t=1,\forall \{l,k\}\in \{l^*,k^*\}$\\}
\end{algorithm}

\vspace{-2mm}
\subsection{Benchmark and no-mirror scenario}
To the best of the authors' knowledge, this is the first work that studies the outage probability performance in an indoor mirror- or ORIS-assisted VLC scenario. As introduced in Section\,\ref{Intro}, prior mirror- or ORIS-assisted VLC works focused on SINR, data rate or secrecy rate maximization. It is thus difficult to compare our proposal to previous works.

As our benchmark, we consider a typical room whose lighting infrastructure satisfies the required illumination conditions in an energy-efficient manner, i.e., minimum total optical power, which can be formulated as 
\vspace{-2mm}
\begin{IEEEeqnarray}{lCl}
\vspace{-1mm}
\label{Benchmark1}
\mathbf{P}^*& = &{\mathop {\argmin }\limits_{{\mathbf{P}},{E_{\rm min}}} \sum\limits_l {P_l} }\\
&&{\rm{subject\,to}}\,{\rm{C5-8, C10, C13}}.\IEEEnonumber
\vspace{-2mm}
\end{IEEEeqnarray}
Then, the room is provided with mirrors (or ORISs) that contribute to minimizing the outage probability. 
This is defined by the variable $\boldsymbol{\beta}$ and computed with the heuristic MM approach to solving subroutine 1  represented in Algorithm\,\ref{alg:Subroutine1MM}. For the benchmark, this process is not iterated (non-alternating), accepting these one-shot values as final. 

We also compare our results with a scenario without mirrors (neither stationary nor ORIS), and the optical power allocated is computed by \eqref{Benchmark1}.

\vspace{-4mm}
\subsection{Complexity analysis}
As mentioned earlier, the \textsc{JointMinOut} optimization problem is a variant of a MIP, which is NP-complete and requires exponential time with respect to the input number of variables to be solved. There is no algorithm able to solve it in polynomial time, and thus, it is intractable to solve in a reasonable time. 
We propose two heuristic approaches whose complexity orders are much lower and analyzed as follows. Table\,\ref{tab:ComplexityAlgorithms} includes a summary of the complexity for each of the algorithms under evaluation in this paper.

\setlength{\tabcolsep}{1pt}
\begin{table}[t]
    \centering
     \caption{Complexity of the algorithms proposed. }
    \begin{tabular}{|c|c|c|}
    \cline{2-3}
     \multicolumn{1}{c|}{} & \textbf{Algorithm} & \textbf{Complexity} \\
    \hline
      Optimal & \textsc{JointMinOut} & Exponential: MIP\\
          \hline
      \multirow{4}{*}{Heuristics} & MM & Polynomial: $O(T_{{\rm max}}N_{{\rm max}}L^2K^2)$ \\
       & MP & Polynomial: $O(T_{{\rm max}}L^2K^2)$ \\
       & Benchmark & Polynomial: $O(N_{{\rm max}}L^2K^2)$\\
       & No mirror & Polynomial: $O\big((L^2+L+1)^2(2n+L+3)\big)$\\
    \hline
        \end{tabular}
    \label{tab:ComplexityAlgorithms}
    \vspace{-4mm}
\end{table}
\setlength{\tabcolsep}{6pt}

Both MM and MP approaches are alternating algorithms. Let us obtain the worst-case complexity order by assuming that the number of iterations in the proposed generalized alternating algorithm is $T_{{\rm max}}$. The MM approach to subroutine\,1  (Algorithm\,\ref{alg:Subroutine1MM}) carries out $N_{{\rm max}}$ times a \emph{selection sort algorithm}, leading to a complexity order of $O(N_{{\rm max}}L^2K^2)$. Differently, the MP approach to subroutine\,1 (Algorithm\,\ref{alg:Subroutine1MP}) carries out a single \emph{selection sort algorithm}, leading to a complexity order of $O(L^2K^2)$. In both approaches, subroutine\,2 is a linear programming problem whose complexity order can be approximated in practice as $O(v^2c)$, where $v$ is the number of variables and $c$ is the number of constraints~\cite{boyd_vandenberghe_2004}.  After converting quadratic into linear variables, there are $v=L^2+L+2$ variables and $c=2n+L+5$ constraints in subroutine\,2. Since the number of wall elements is much larger than the number of LEDs $(K>>L)$, the complexity of subroutine\,1 exceeds that of subroutine 2, and the global complexity order can be approximated as $O(T_{{\rm max}}N_{{\rm max}}L^2K^2)$ for the MM approach and $O(T_{{\rm max}}L^2K^2)$ for the MP approach.

The benchmark algorithm is a linear programming solution similar to subroutine 2, followed by the MM heuristic approach (subroutine 1 - Algorithm\,\ref{alg:Subroutine1MM}). As stated above, the complexity of subroutine 1 exceeds that of subroutine 2, and therefore the complexity of the benchmark algorithm can be approximated as $O(N_{{\rm max}}L^2K^2)$. The no-mirror algorithm is computed only by \eqref{Benchmark1}, which is a linear programming problem with a complexity of $O\left((L^2+L+1)^2(2n+L+3)\right)$.

\section{Results and Discussion}\label{ResultsAndDiscussion}

In this section, we present comprehensive simulation results for the proposed mirror- or ORIS-aided VLC system. Without loss of generality, we consider a residential room of 4\,x\,4\,x\,3\,m, with a total of $L=4$ LEDs deployed in a symmetric 2-by-2 lattice with coordinates [1, 1], [1, 3], [3, 1] and [3, 3]\,m. Though we consider NLoS contributions from the four walls, only one wall is provided with mirror (or ORIS) capabilities. Each wall is divided into $K=K_{\rm{x}}\times K_{\rm{y}}=30\times15=450$ elements. 

Due to the deterministic characteristics of the VLC channel, we assume an object model for the link-path blockage~\cite{CylinderBlockage}. The proposed model considers only self-blocking, i.e., a situation of an empty room where blockages of LoS and NLoS paths are produced by the body of the person holding the user device. However, this study could be easily extended to other scenarios with static objects such as pillars or furniture, which are even easier to model due to their immobility, or with other users moving around, which typically produce less probable blockage. Specifically, in this paper, the human body is modeled by a cylinder of height 1.75\,m and radius 0.15\,m, holding a device provided with a single PD looking upwards and separated at a distance of 0.3\,m from the body, with a horizontal angle following a uniform distribution $\mathcal{U}[0, 2\pi)$~\cite{CylinderBlockage}. 

We assume an LED with a typical luminous efficacy of 280\,lm/W (theoretical values for white LEDs can be above 300\,lm/W~\cite{LuminousEfficacy}), and we consider a scenario where the required minimum average illuminance is 500\,lux, with a maximum illuminance allowed at each point of 800\,lux, and a minimum illuminance uniformity of 0.5. Note that these are the minimum illumination conditions for a typical office scenario in which writing, typing or reading tasks are carried out~\cite{LightingStandard}. As argued in Section\,\ref{subsec:MirrorVSRIS}, because the NLoS path loss will not be heavily affected by the $\phi_{1/2}$ parameter considering the presented scenario, we select the $\phi_{1/2}$ value to preferably satisfy illumination requirements, i.e., wide enough to support the lighting uniformity. More detailed simulation parameters are shown in Table\,\ref{tab:SystemParameters}. 

\setlength{\tabcolsep}{4pt}
\begin{table}[!t]
\vspace{0.05in}
\caption{Simulation parameters.}
\vspace{-2mm}
\label{tab:SystemParameters}
\centering
{\footnotesize
\begin{tabular}{llcr}
\hline
\hline
Notation & Parameter description & Value & Unit\\
\hline
- & LED height &  3 & [m]\\
- & User device height &  1 & [m]\\
$\phi_{1/2}$ & \makecell[l]{Half-power semi-angle of \\ the LED} &  80 & [deg.]\\
$\tilde{r}$ & Reflection coefficient of wall & 0.2 & [-]\\
$\hat{r}$ & Reflection coefficient of mirror/ORIS & 0.99 & [-]\\
$\rho$ & PD responsivity & 1 & [A/W]\\
$A_{\mathrm{PD}}$ & PD physical area & 1 & [cm$^2$]\\
$\Psi$ & FoV semi-angle of the PD & [30, 40, 50] & [deg.]\\
$B$ & Communication bandwidth & 20 & [MHz]\\
$N_0$ & \makecell[l]{Power spectral density of\\ the AWGN} & $2.5{\cdot}10^{-20}$ & [W/Hz]\\
$K_{\rm e/v}$ & Luminous efficacy of LEDs & 280 & [lm/W]\\
$\gamma_{\rm th}$ & SNR threshold & [10 - 50] & [dB]\\
$N_{\rm max}$ & \makecell[l]{Maximum number of mirrors\\ or ORIS elements} & 128 & [-]\\
$E_{\rm th}$ & \makecell[l]{Minimum average illuminance \\required} & 500 & [lux]\\
$E_{\rm max}$ & \makecell[l]{Maximum illuminance allowed \\at each point} & 800 & [lux]\\
$U_{\rm min}$ & Illuminance uniformity & 0.5 & [-]\\
$T_{\rm max}$ & \makecell[l]{Maximum number of iterations\\ of the AO algorithm} & 20 & [-]\\
\hline
\hline
\end{tabular}}
\vspace{-5mm}
\end{table}
\setlength{\tabcolsep}{6pt}

Note that the communication bandwidth and PD physical area selected for our simulations are 20\,MHz and 1\,cm$^2$, respectively, which are within the ranges of actual systems~\cite{DLPerformanceAttocell} and aligned with the principles of ultra-reliable non-rate-centric broadband communication for the Internet of Everything~\cite{AVisionOf6G}. Besides, note that the simulation parameters allow us to work in a small detector regime that, together with our LED point-source assumption and the neglected inter-element blockage, allows us to apply the specular reflection channel model formulated in \eqref{eq:ChannelModelNLoSmirror} and \eqref{eq:ChannelModelNLoSRIS} for mirror and ORIS cases, respectively.



\begin{figure}[t]
\centering
\includegraphics[width=\columnwidth]{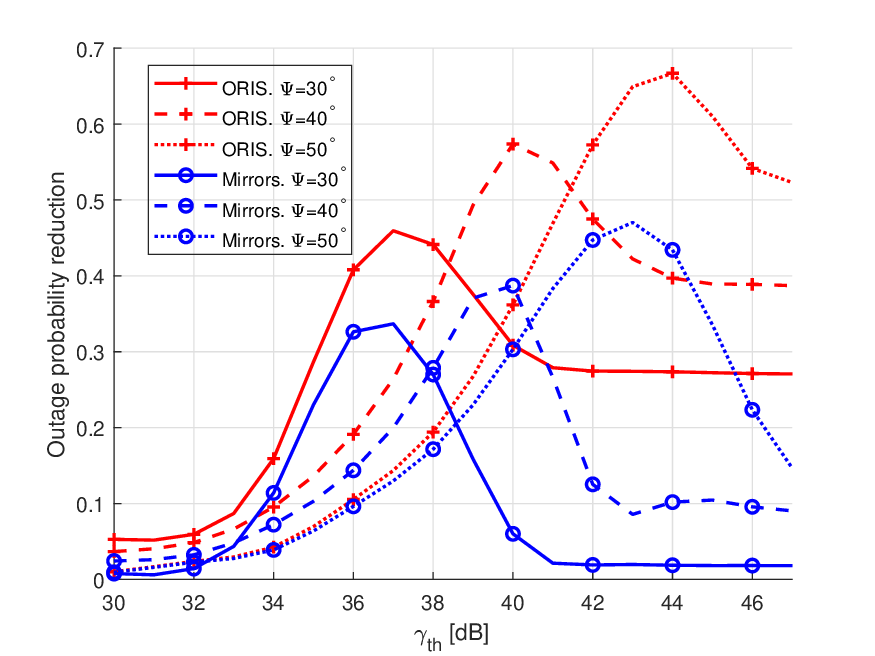}
        \vspace{-6mm}\caption{Outage probability reductions obtained with MP approach when using ORIS (i.e. mirrors with mobile orientation) or mirror elements with respect to a no-mirror scenario for different $\Psi$ angles.}
\label{fig:OutageProbGain}
\vspace{-5mm}
\end{figure}

\begin{figure}[t]
     \centering
     \begin{subfigure}[t]{0.32\columnwidth}
         \centering
         \includegraphics[width=\textwidth]{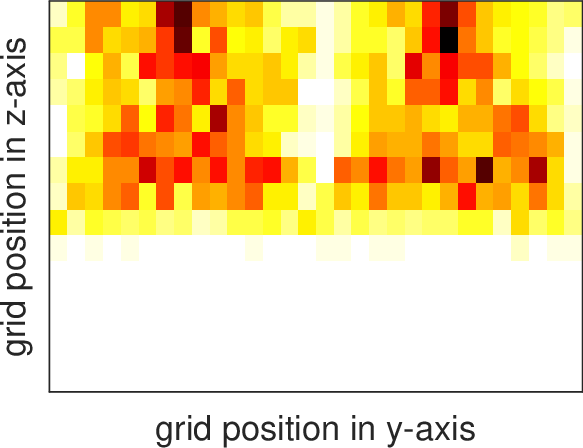}
         \caption{$\Psi=30^\circ$}
         \label{fig:RIS_MM_FoV30}
     \end{subfigure}
     \hfill
     \begin{subfigure}[t]{0.32\columnwidth}
         \centering
         \includegraphics[width=\textwidth]{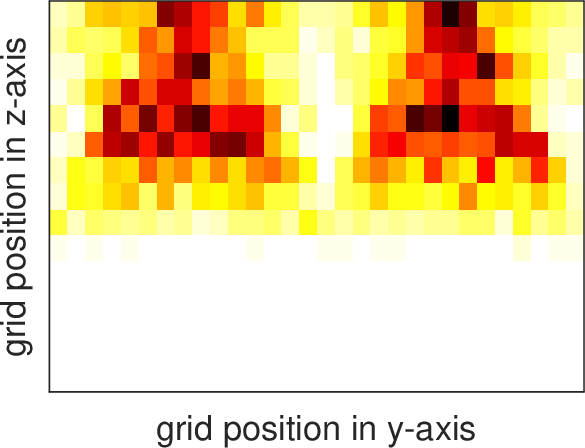}
         \caption{$\Psi=40^\circ$}
         \label{fig:RIS_MM_FoV40}
     \end{subfigure}
     \hfill
     \begin{subfigure}[t]{0.32\columnwidth}
         \centering
         \includegraphics[width=\textwidth]{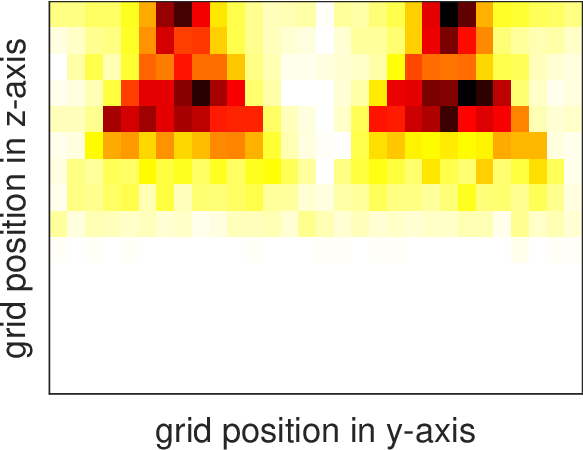}
         \caption{$\Psi=50^\circ$}
         \label{fig:RIS_MM_FoV50}
     \end{subfigure}
     \vspace{-2mm}
        \caption{Heatmaps of optimal ORIS placement on the wall for MM approach when $\gamma_{\rm th}$=40\,dB at different FoV semi-angles. Darker pixels indicate a higher probability.}
         \label{fig:RIS_MM}
         \vspace{-4mm}
\bigbreak
     \centering
     \begin{subfigure}[t]{0.32\columnwidth}
         \centering
         \includegraphics[width=\textwidth]{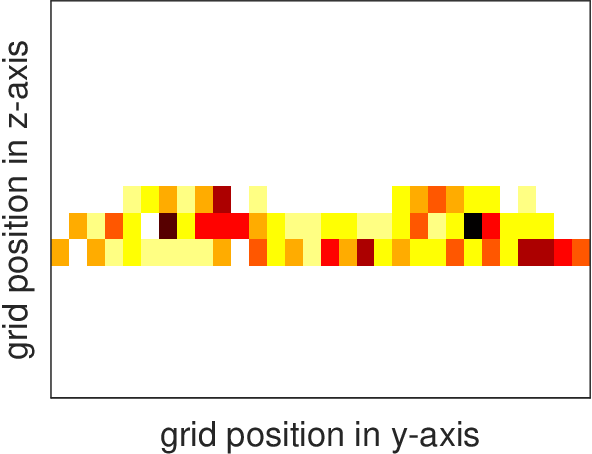}
         \caption{$\Psi=30^\circ$}
         \label{fig:Mirror_MM_FoV30}
     \end{subfigure}
     \hfill
     \begin{subfigure}[t]{0.32\columnwidth}
         \centering
         \includegraphics[width=\textwidth]{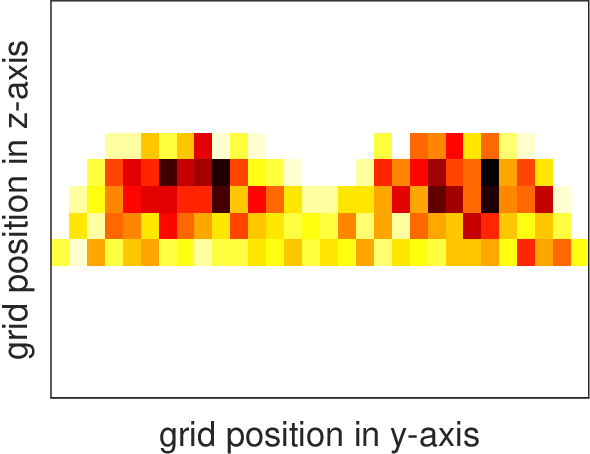}
         \caption{$\Psi=40^\circ$}
         \label{fig:Mirror_MM_FoV40}
     \end{subfigure}
     \hfill
     \begin{subfigure}[t]{0.32\columnwidth}
         \centering
         \includegraphics[width=\textwidth]{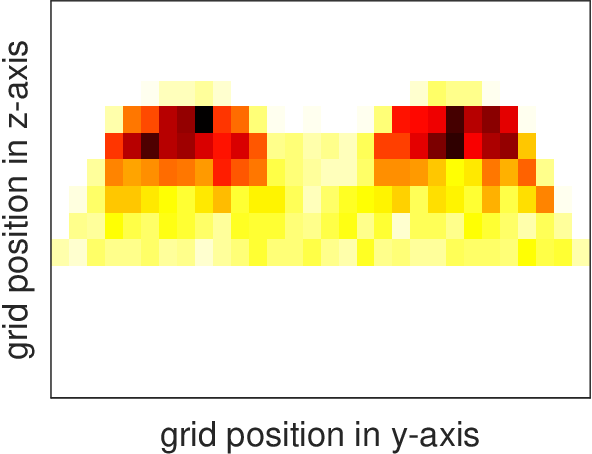}
         \caption{$\Psi=50^\circ$}
         \label{fig:Mirror_MM_FoV50}
     \end{subfigure}
     \vspace{-2mm}
        \caption{Heatmaps of optimal mirror placement on the wall for MM approach when $\gamma_{\rm th}$=40\,dB at different FoV semi-angles. Darker pixels indicate a higher probability.}
         \label{fig:Mirror_MM}
         \vspace{-4mm}
\bigbreak
     \centering
     \begin{subfigure}[t]{0.32\columnwidth}
         \centering
         \includegraphics[width=\textwidth]{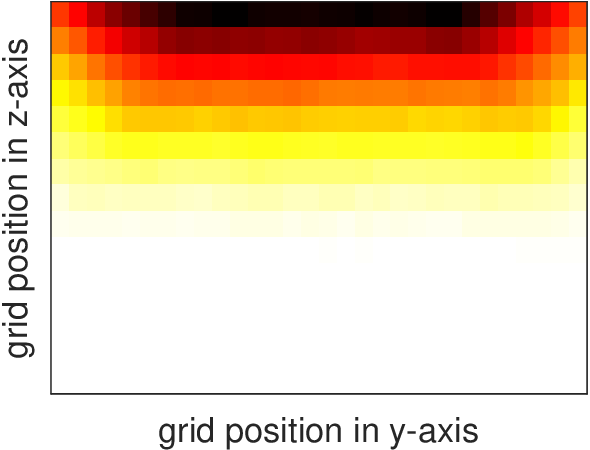}
         \caption{$\Psi=30^\circ$}
         \label{fig:RIS_MP_FoV30}
     \end{subfigure}
     \hfill
     \begin{subfigure}[t]{0.32\columnwidth}
         \centering
         \includegraphics[width=\textwidth]{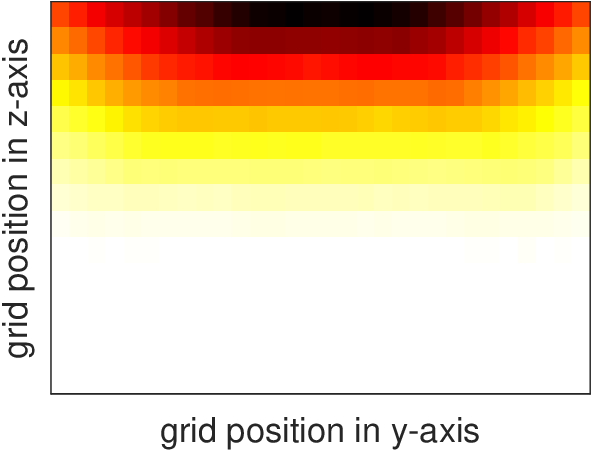}
         \caption{$\Psi=40^\circ$}
         \label{fig:RIS_MP_FoV40}
     \end{subfigure}
     \hfill
     \begin{subfigure}[t]{0.32\columnwidth}
         \centering
         \includegraphics[width=\textwidth]{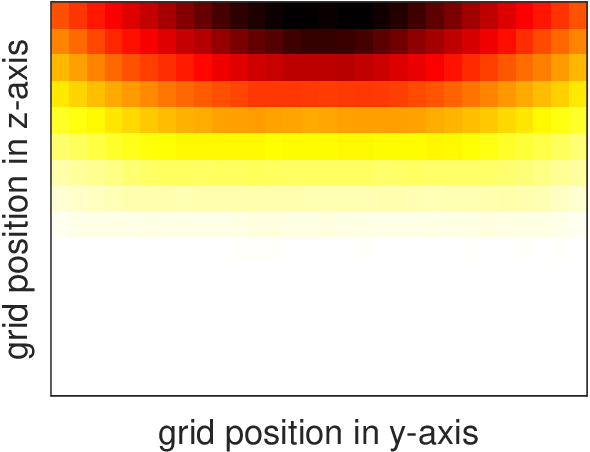}
         \caption{$\Psi=50^\circ$}
         \label{fig:RIS_MP_FoV50}
     \end{subfigure}
     \vspace{-2mm}
        \caption{Heatmaps of optimal ORIS placement on the wall for MP approach when $\gamma_{\rm th}$=40\,dB at different FoV semi-angles. Darker pixels indicate a higher probability.}
         \label{fig:RIS_MP}
         \vspace{-4mm}
\bigbreak
     \centering
     \begin{subfigure}[t]{0.32\columnwidth}
         \centering
         \includegraphics[width=\textwidth]{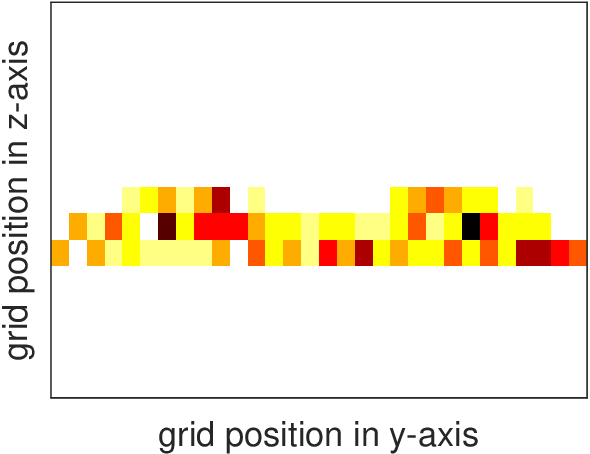}
         \caption{$\Psi=30^\circ$}
         \label{fig:Mirror_MP_FoV30}
     \end{subfigure}
     \hfill
     \begin{subfigure}[t]{0.32\columnwidth}
         \centering
         \includegraphics[width=\textwidth]{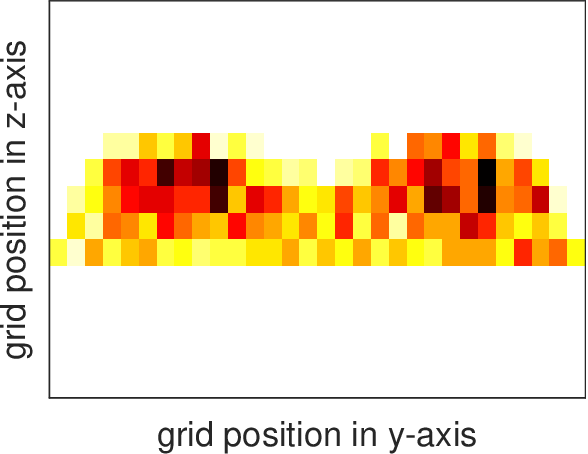}
         \caption{$\Psi=40^\circ$}
         \label{fig:Mirror_MP_FoV40}
     \end{subfigure}
     \hfill
     \begin{subfigure}[t]{0.32\columnwidth}
         \centering
         \includegraphics[width=\textwidth]{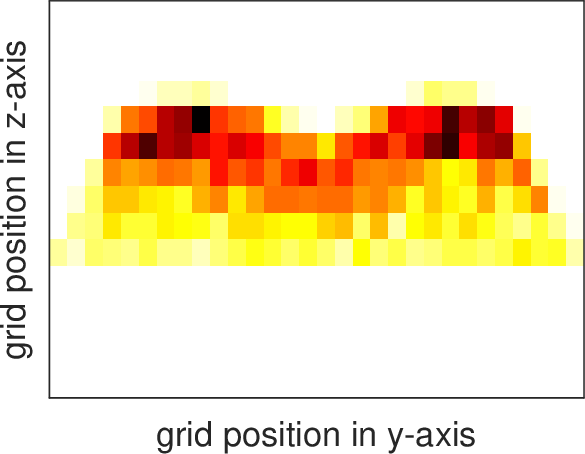}
         \caption{$\Psi=50^\circ$}
         \label{fig:Mirror_MP_FoV50}
     \end{subfigure}
     \vspace{-2mm}
        \caption{Heatmaps of optimal mirror placement on the wall for MP approach when $\gamma_{\rm th}$=40\,dB at different FoV semi-angles. Darker pixels indicate a higher probability.}
         \label{fig:Mirror_MP}
         \vspace{-7mm}
\end{figure}

\vspace{-4mm}
\subsection{Comparison between using mirrors, ORISs, and no mirrors}
Let us consider a scenario where either mirrors or ORISs are deployed optimally to support each realization of the user's position. We run the proposed AO algorithm for the MP approach to see the contribution of mirrors or ORISs and compare their performance with a scenario where there are no mirrors. Fig.\,\ref{fig:OutageProbGain} plots the outage probability reduction obtained with respect to a scenario without mirrors. The use of ORIS elements allows an outage reduction of up to 0.67, 0.58 and 0.46 when $\Psi$=50$^\circ$, $\Psi$=40$^\circ$ and $\Psi$=30$^\circ$, respectively. Mirrors, due to their immobility, provide lower outage reductions of about 0.48, 0.39 and 0.33 for $\Psi$=50$^\circ$, $\Psi$=40$^\circ$ and $\Psi$=30$^\circ$, respectively. Note that there is a higher gap among the maximums of the outage probability reduction curves in the case of ORIS as compared to the case of mirrors; this is because the mobility of ORIS gives them one more degree of freedom to enhance the NLOS path and, as a consequence, to reduce the outage probability at a larger extent when the FoV angle increases. Thus, whenever possible, ORIS should be deployed. However, when the extra complexity of ORIS deployment due to its mobile orientation is not affordable, the deployment of mirrors will also provide a considerable outage reduction.

\begin{figure}[t]
\centering
\includegraphics[width=0.95\columnwidth]{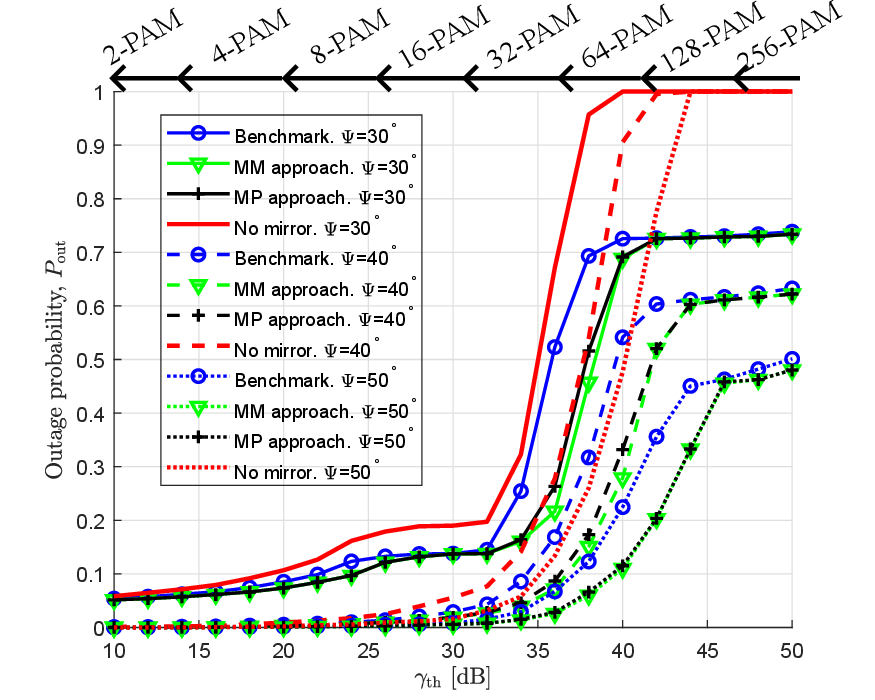}
        \caption{Outage probability vs. $\gamma_{\rm th}$ for the benchmark, no-mirror case, and MM and MP proposed algorithms using different $\Psi$.}
\label{fig:OutageProb_vs_SNRth_RIS}
\vspace{-4mm}
\end{figure}

\vspace{-4mm}
\subsection{Optimal placement of mirrors}
After evaluating the maximum gain obtained when either ORISs or mirrors are deployed, we analyze the optimal fixed placement for these elements in each of the suggested approaches. We consider a $\gamma_{\rm th}=40$\,dB. Let us first analyze the optimal placement of ORISs and mirrors with the MM approach, represented in Figs.~\ref{fig:RIS_MM} and~\ref{fig:Mirror_MM}, respectively, for three different $\Psi$ angles. The figures reflect the placements of mirrors/ORISs that contribute to closing the SNR gap the most often. 
Note that, in the case of ORIS, the optimal  placement is high on the wall, while in the case of mirrors, their lack of mobility imposes an optimal placement halfway up the wall. 
Also, note that the optimal reflector position is much clearer at higher $\Psi$ values where the contributions from NLoS are more important (see Section\,\ref{subsec:MirrorVSRIS}).

The optimal placement of ORISs and mirrors when invoking the MP approach is represented in Figs.~\ref{fig:RIS_MP} and~\ref{fig:Mirror_MP}, respectively. Again, darker pixels indicate a higher probability. Unlike the MM approach, the MP approach exploits ORISs (or mirrors) as much as possible, and then the shape of the optimal placement zone is not as well defined as when invoking the MM approach. The optimal placement of ORIS for the MP approach is in the upper edge of the wall; in any room, ORIS elements so installed provide much improvement in outage probability. In fact, ORIS located at such a high level on the wall is unlikely to impact the utilization of the space, i.e., other objects or decorations can still be placed in the middle of the wall, as they normally would be. 

Due to the benefits of using ORISs compared to mirrors, we consider only ORIS elements for the results in the following sections.

\vspace{-4mm}
\subsection{Comparison between the benchmark, no-mirror, and the MM and MP proposed algorithms}

First, we compare the performance obtained with the MM and MP proposed approaches to the non-alternating benchmark and a no-mirror scenario for a range of $\gamma_{\rm th}$ values from 10\,dB (low demanding wireless services) to 50\,dB (high demanding wireless services), shown in Fig.\,\ref{fig:OutageProb_vs_SNRth_RIS}. The SNR threshold range includes all the values required to demodulate  $2$-PAM through $256$-PAM at a bit error rate (BER) equal to 10$^{-6}$ in an IM/DD system~\cite{OWCMatlab}. These results are depicted for three different $\Psi$ values. The smaller the $\Psi$, the higher the outage probability is because less contribution is received from neighboring LEDs as well as from reflecting elements. In fact, for $\Psi$=30$^\circ$, there is an outage probability floor, which means that there are some room areas that cannot be reached by any light source and where the user will be in outage. A zero-outage probability is obtained for larger $\Psi$ values and $\gamma_{\rm th}<25$\,dB. When comparing the three approaches, both MM and MP provide similar results in terms of outage probability, considerably improving upon the benchmark and no-mirror scenarios. As an example, for $\gamma_{\rm th}=40$\,dB and $\Psi$=50$^\circ$, both proposed approaches, MM and MP, reduce the outage probability by two and five times with respect to the benchmark and no-mirror scenarios, respectively. This means that, for the same outage probability performance, the proposed AO algorithms enable more demanding services requiring high $\gamma_{\rm th}$ values. 

\begin{figure}[t]
\centering
\includegraphics[width=\columnwidth]{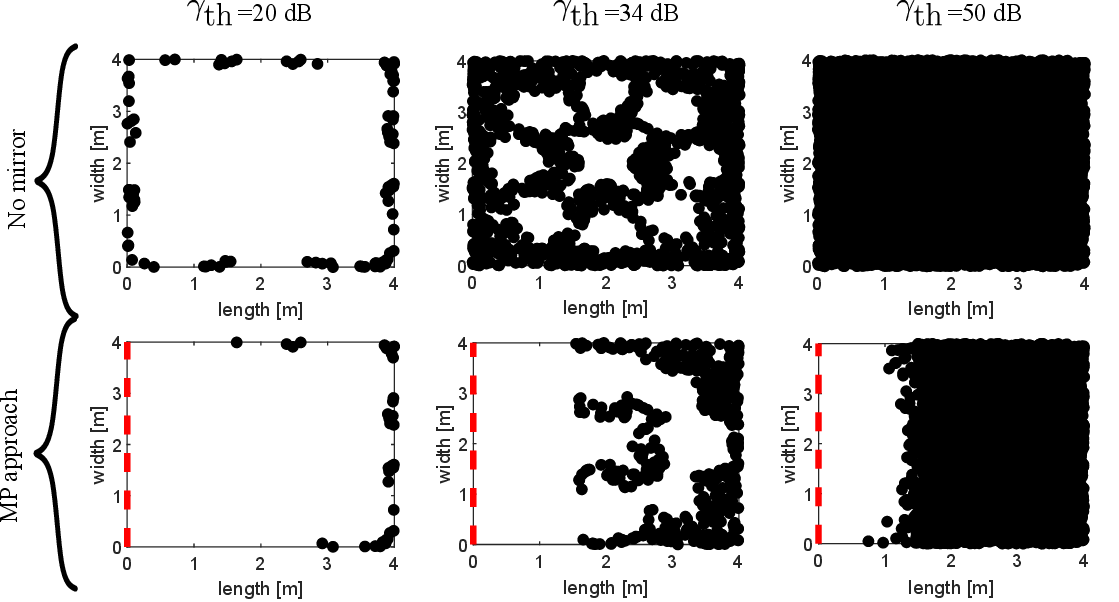}
        \caption{Top view of the room showing, using black circles, the location of users ($\Psi=40^\circ$) in outage for different $\gamma_{\rm th}$ values for a Monte Carlo simulation of $10^4$ trials. 
        The no-mirror (first row) and ORISs using MP (second row) approaches are invoked. The red dashed line on the left wall represents where ORISs are deployed.}
\label{fig:MapOutage}
\vspace{-3mm}
\end{figure}

\begin{figure}[t]
\centering
\includegraphics[width=\columnwidth]{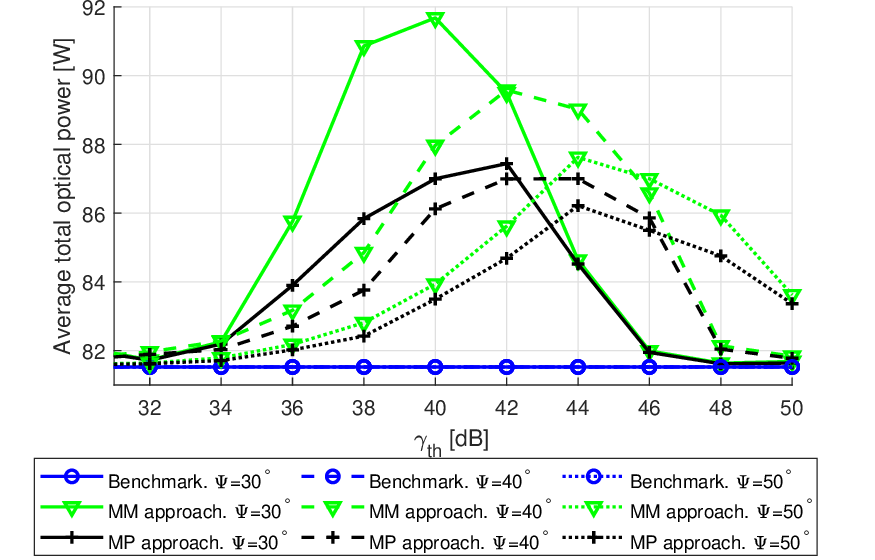}
\vspace{-6mm}
        \caption{Average total optical power vs. $\gamma_{\rm th}$ for the benchmark, MM, and MP proposed algorithms using different $\Psi$.}
\label{fig:AveragePower_vs_SNRth_RIS}
\vspace{-3mm}
\end{figure}

Fig.\,\ref{fig:MapOutage} shows realizations for the location of users that are in outage assuming  $\gamma_{{\rm th}}$=20\,dB, $\gamma_{{\rm th}}$=34\,dB or $\gamma_{{\rm th}}$=50\,dB for the no-mirror and MP approaches, for a simulation of $10^4$ trials. The results clearly show how the MP approach supports those users located close to the wall provided with ORIS elements, even at high $\gamma_{{\rm th}}$ values. In contrast, if no mirror is provided, the entire area can easily be in outage. Note that if all walls were provided with ORIS elements, the outage probability would dramatically decrease to almost zero, at the expense of a higher complexity and coordination among ORIS elements located on different walls.

\begin{figure}[t]
\centering
\includegraphics[width=\columnwidth]{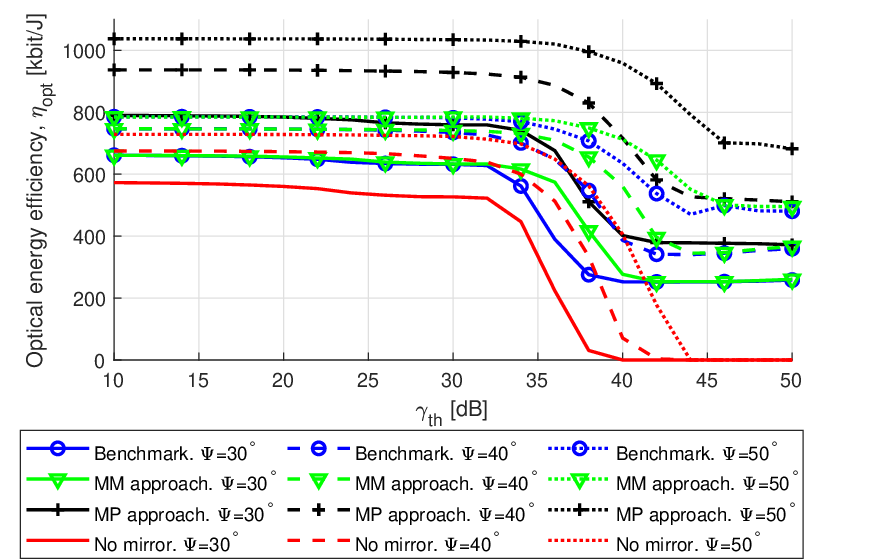}
        \caption{Optical energy efficiency vs. $\gamma_{\rm th}$ for benchmark, no-mirror case, and MM and MP proposed algorithms when using different $\Psi$.}
\label{fig:SpectralEfficiency_vs_SNRth_RIS}
\vspace{-5mm}
\end{figure}

\vspace{-2mm}
\subsection{Analyzing resources used}
Let us now evaluate the two available resources to conserve: total power $\left(\sum_lP_l\right)$, and the number of ORIS elements $\left(\sum_l\sum_k\beta_{l,k}\right)$. Fig.\,\ref{fig:AveragePower_vs_SNRth_RIS} plots the average total optical power as a function of $\gamma_{\rm th}$. 
Though both MM and MP provide a similar outage probability performance, the MM approach achieves this performance by increasing the transmitted power. 
Both approaches increase the transmitted power when $\gamma_{\rm th}$ becomes more demanding, but decrease it when it becomes so large that the user can no longer be supported at all. The benchmark keeps the total optical power constant regardless of the $\gamma_{\rm th}$ value; this is because the variables are computed as one shot optimizations of $\mathbf{P}$ and $\boldsymbol{\beta}$ independent from each other. Note that the power consumption in a no-mirror scenario is the same as for the benchmark. 

Results shown in Fig.\,\ref{fig:OutageProb_vs_SNRth_RIS} and Fig.\,\ref{fig:AveragePower_vs_SNRth_RIS} are combined to compute the optical energy efficiency formulated in \eqref{eq:SpecEfficiency}. Fig.\,\ref{fig:SpectralEfficiency_vs_SNRth_RIS} shows the optical energy efficiency as the required $\gamma_{\rm th}$ varies. The MP approach is much more efficient than any other approach, and this difference increases for larger $\Psi$ values. 

\begin{figure}[t]
\centering
\includegraphics[width=\columnwidth]{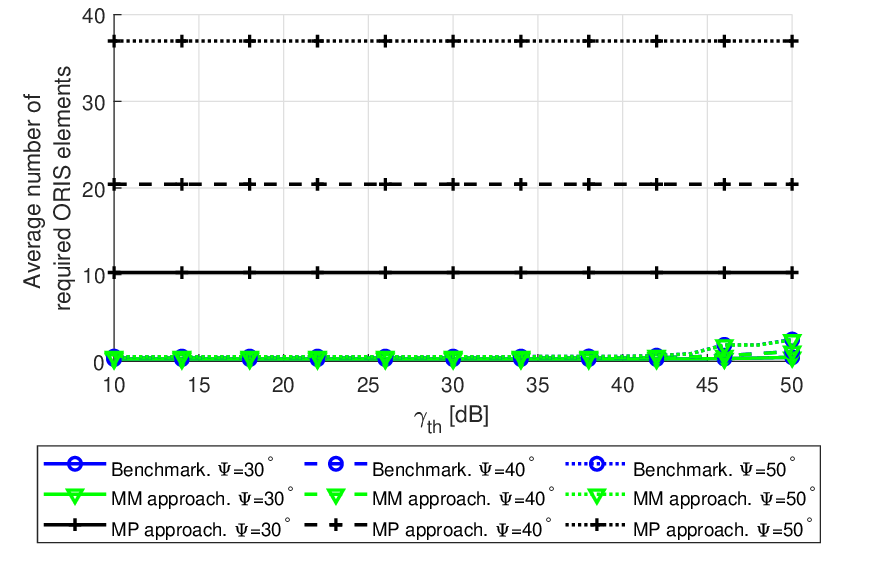}
        \caption{Number of ORIS elements required vs. $\gamma_{\rm th}$ for the benchmark, MM, and MP proposed algorithms using different $\Psi$.}
\label{fig:NrisRequired_vs_SNRth_RIS}
\vspace{-2mm}
\end{figure}

We now evaluate the number of ORIS elements required to achieve the outage probability obtained in Fig.\,\ref{fig:OutageProb_vs_SNRth_RIS}. This is represented in Fig.\,\ref{fig:NrisRequired_vs_SNRth_RIS} for the entire $\gamma_{\rm th}$ range. The advantage of the MP approach in optical energy efficiency comes at a cost of requiring a much larger number of ORIS elements. The larger the $\Psi$, the more ORIS elements can be exploited as the user may receive larger NLoS contributions. Conversely, the MM and benchmark algorithms only exploit ORIS elements when it is strictly necessary, i.e., at high $\gamma_{\rm th}$ values when they cannot allocate more transmitted power per LED to support users due to the illumination constraints.

\begin{table}
  \setlength{\tabcolsep}{2pt} 
\renewcommand{\arraystretch}{1.5} 
  \caption{Probability of the number of iterations required for convergence of the proposed MP and MM alternating iterative algorithms over a range of $\gamma_{\rm th}$ values, assuming ORIS elements are deployed and $\Psi=50^\circ$.}
  \label{tab:Niter}
  \subfloat[MP approach]{%
    \label{tab:NiterMP}%
    \begin{tabular}{c|c|c}
                                              Iterations & ${\le} 4$  & $T_{\rm max}$  \\ \hline 
                     $\gamma_{\rm th}{=}[10,28]$\,dB                    & 100.00\% & 0.00\%   \\ \hline 
                            $\gamma_{\rm th}{=}(28,50]$\,dB                    & 99.69\%  & 0.31\%  
    \end{tabular}}
  \hfill%
   \subfloat[MM approach]{%
    \label{tab:NiterMM}%
    \begin{tabular}{c|c|c}
                                               Iterations & $\le 4$  & $T_{\rm max}$  \\ \hline 
                     $\gamma_{\rm th}{=}[10,24]$\,dB                    & 100.00\% & 0.00\%   \\ \hline 
                            $\gamma_{\rm th}{=}(24,50]$\,dB                    & 99.73\%  & 0.26\%  
    \end{tabular}}
    \vspace{-3mm}
\end{table}

\vspace{-3mm}
\subsection{Convergence time for the AO algorithms proposed}
The number of iterations for convergence of the proposed AO algorithm is represented in Table\,\ref{tab:Niter} for all $\gamma_{\rm th}$ values considered. Both MM and MP approaches need less than 5 iterations to converge, which is encouraging for a real deployment. Note that there are few user positions that need a large number of iterations, i.e., they do not converge and reach $T_{\rm max}=20$ in the AO algorithm. However, these users are rare ($<0.50\%$) and do not affect the global system performance.


\vspace{-3mm}
\section{Conclusion}
\label{sec:Conclusion}
In this paper, we formulated an optimization problem called \textsc{JointMinOut} to minimize the outage probability in a room with multiple LEDs, where a wall can be provided either with mirrors or with ORIS elements, and where the user body is considered as possibly blocking the LoS and NLoS links. A range of required SNRs is evaluated subject to multiple standard illumination constraints imposed to guarantee the dual functionality of VLC. We have geometrically analyzed the capability of ORISs and mirrors for supporting users. Since the optimization problem is intractable due to being NP-complete, we proposed two heuristic iterative approaches and compared their performance to a benchmark and a no-mirror scenario. Numerical results show that the proposed approaches provide reductions in the outage probability of up to 67\%. Besides, the convergence of the proposed approaches is analyzed, showing that a small number of iterations is required. We have also studied the optimal placement of ORISs and mirrors for each proposed approach, and our results determined that an ORIS-aided VLC scenario invoking the MP approach provides considerable gains in outage probability and energy efficiency, while the optimal ORIS deployment is high on the wall and does not affect the room's usage because they are not placed at a standing or sitting human eye level.






\vspace{-3mm}
\bibliography{./references}
\bibliographystyle{IEEEtran}


\section{Biography Section}
 
\vspace{-1.2cm}

\begin{IEEEbiography}[{\includegraphics[width=1in,height=1.25in,clip,keepaspectratio]{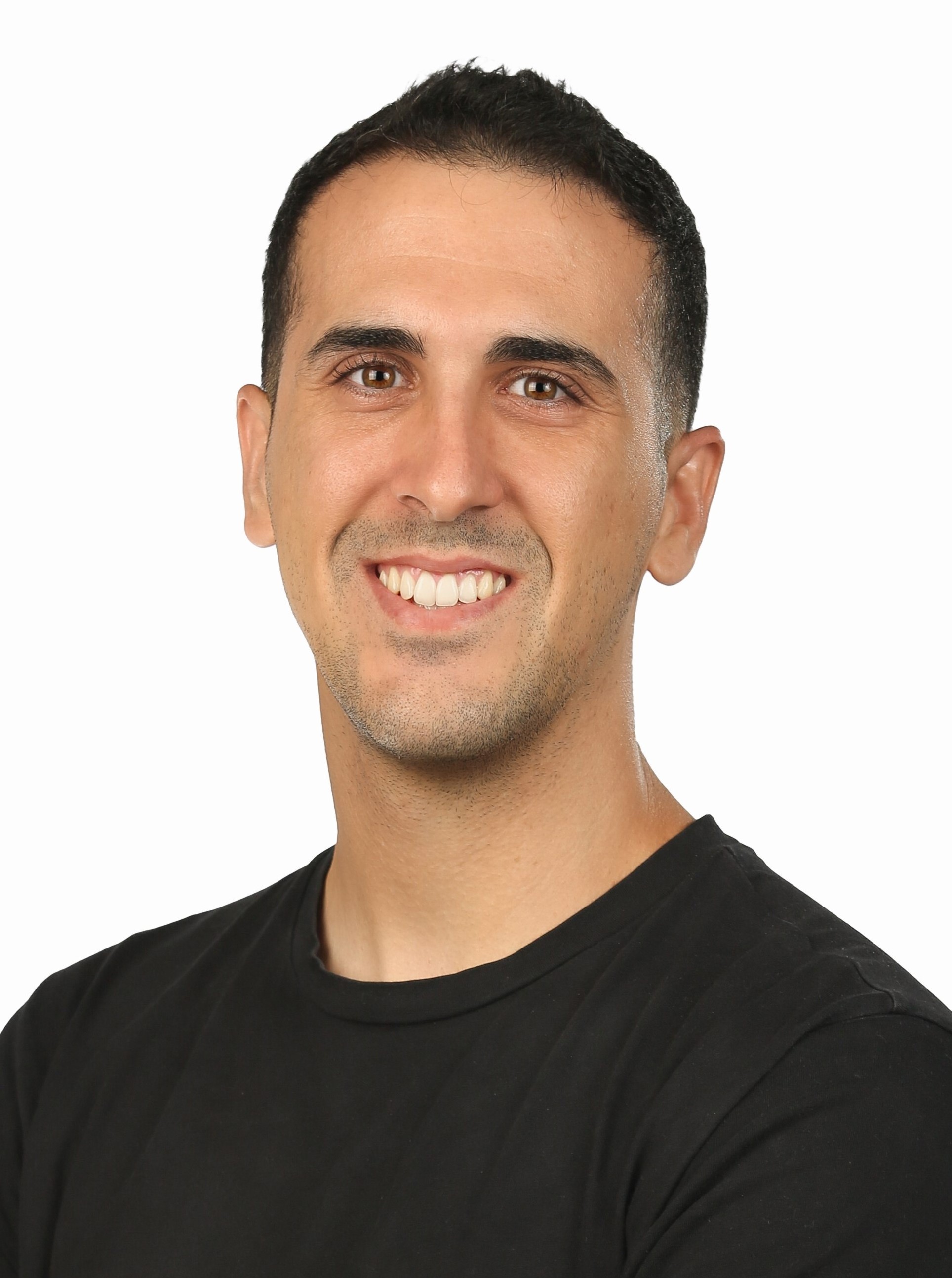}}]{Borja Genoves Guzman} (Senior Member, IEEE) received a B.Sc. degree from Universidad Carlos III de Madrid, Spain, in 2013, a double M.Sc. degree (Hons.) in electrical engineering from Universidad Carlos III de Madrid, and the Institut Mines-Télécom, France, in 2015, and a Ph.D. degree from University Carlos III of Madrid in 2019 (Extraordinary Ph.D. award). He received the First Prize in Graduation National Awards by the Ministry of Education, Culture and Sports from Spain. He worked as a postdoctoral researcher for 3 years at IMDEA Networks Institute. He has carried out research stays at University of Virginia (USA) - 2 years, University of Southampton (UK) - 3 months and The University of Edinburgh (UK) - 3 months. He is currently working as a Marie Skłodowska-Curie Actions Postdoctoral Global Fellow in Universidad Carlos III de Madrid, and starting as a Ramón y Cajal researcher from September 2025. He has published over 45 peer-reviewed contributions. He also holds 2 patents (1 submitted and 1 granted). His research career has been supported by 3 regional, 3 national, and 5 European research projects, 2 of them as Principal Investigator. Since the beginning of his research career, he has obtained funding for his personal research projects: Erasmus+ and FPU in the pre-doctoral career stage; and Juan de la Cierva-Formación, Marie Curie-Postdoctoral Global Fellowship and Ramón y Cajal in the postdoctoral career stage. He is co-founder and board member of Sensory-Fi S.L. (LiFi4Food product), a startup providing IoT solutions for precision agriculture. His research interests are visible light communication (also referred to as LiFi), Internet of Things (IoT) and low-power wireless communications.
\vspace{-1.2cm}

\end{IEEEbiography}
\begin{IEEEbiography}[{\includegraphics[width=1in,height=1.25in,clip,keepaspectratio]{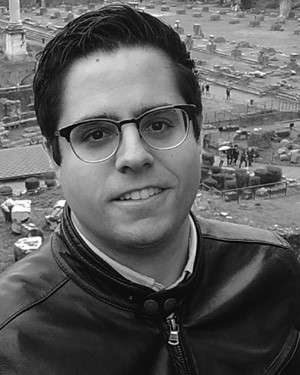}}]{Maximo Morales Cespedes} (Member, IEEE) was born in Valdepeñas, C. Real, Spain, in 1986. He received the B.Sc., and M.Sc., and Ph.D. degrees from the Universidad Carlos III de Madrid, Spain, in 2010, 2012, 2015, respectively, all in electrical engineering, with a specialization in Multimedia and Communications. In 2012 he was finalist of the IEEE Region 8 Student Paper Contest. From 2015 to 2017 he has been working as a postdoctoral fellow with the Institute of Information and Communication Technologies, Electronics and Applied Mathematics (ICTEAM) at Universite Catholique de Louvain. Currently, he is with the Department of Signal Theory and Communications, Universidad Carlos III de Madrid, Spain. He serves in the editorial boards of IEEE Communication Letters and  IEEE Open Journal of the Communications Society. Moreover, he has been TPC member for numerous IEEE conferences (VTC, Globecom, ICC and WCNC). His research interests are interference management, hardware implementations, MIMO techniques and signal processing applied to wireless communications. 
\end{IEEEbiography}
\vspace{-1.2cm}

\begin{IEEEbiography}[{\includegraphics[width=1in,height=1.25in,clip,keepaspectratio]{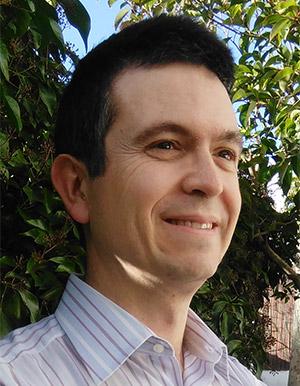}}]{Victor P. Gil Jimenez} (Senior Member, IEEE) received the B.S. degree (Hons.) in telecommunication from the University of Alcal\'a in 1998 and the M.S. degree (Hons.) in telecommunication and the Ph.D. degree (Hons.) from the Universidad Carlos III de Madrid in 2001 and 2005, respectively. He was with the Spanish Antarctica Base in 1999 as a Communications Staff. He visited the University of Leeds, U.K., in 2003, Chalmers Technical University, Sweden, in 2004, and the Instituto de Telecommunicações, Portugal, from 2008 to 2010 and in 2023. He is currently with the Department of Signal Theory and Communications, Universidad Carlos III de Madrid, as an  Associate Professor. He has also led several private and national Spanish projects and has participated in several European and international projects. He holds one patent. He has published over 80 journal articles/conference papers and 9 book chapters. His research interests include advanced multicarrier systems for wireless radio, satellite and visible light communications. He held the IEEE Spanish Communications and Signal Processing Joint Chapter Chair from 2015 to 2023. He received the Master Thesis and the Ph.D. Thesis Award from the  professional Association of Telecommunication Engineers of Spain in 1998 and 2006, respectively. Since 2023, he is Associated Editor at IEEE Open Journal of the Communication Society.
\end{IEEEbiography}
\vspace{-1.2cm}

\begin{IEEEbiography}[{\includegraphics[width=1in,height=1.25in,clip,keepaspectratio]{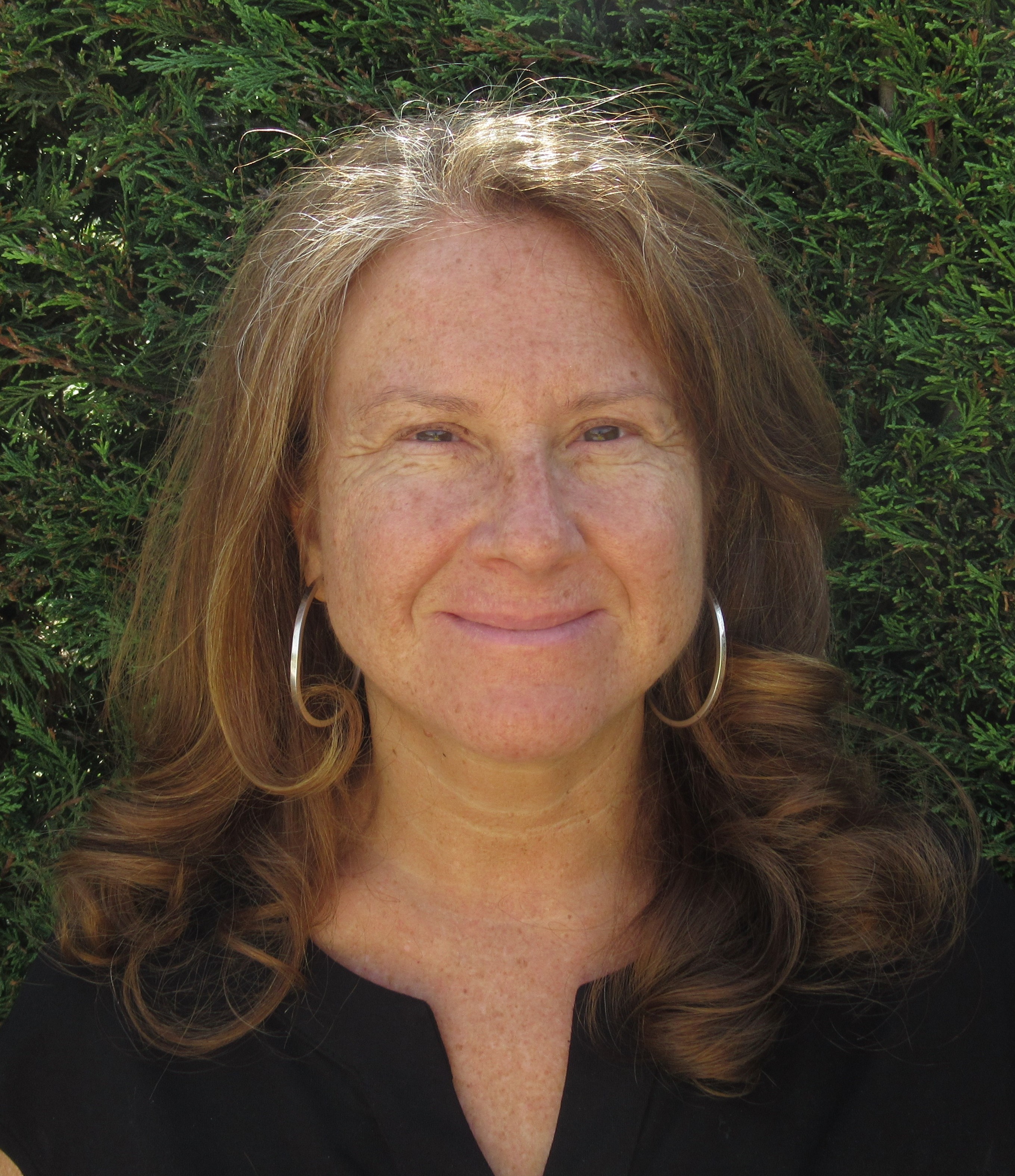}}]{Ana Garcia Armada} (Fellow, IEEE) is a Professor at University Carlos III of Madrid, Spain, where she is leading the Communications Research Group. She has been a visiting scholar at Stanford University, Bell Labs and University of Southampton. She has coordinated a large number of national and international research projects as well as contracts with the industry, all of them related to wireless communications. She has published more than 250 papers in international journals and conference proceedings and she holds seven granted patents. She is serving on the editorial board of IEEE Open Journal of the Communications Society (Associate Editor in Chief since 2024) and ITU Journal on Future and Evolving. She has been a member of the organizing committee of IEEE MeditCom 2024 (General Chair), IEEE MeditCom 2023, IEEE WNCN 2024, IEEE Globecom 2022, IEEE Globecom 2021 (General Chair), IEEE Globecom 2019, IEEE Vehicular Technology Conference (VTC) Fall 2018, Spring 2018 and 2019, among others. She has received the Young Researchers Excellence Award from University Carlos III of Madrid. She was awarded the third place Bell Labs Prize 2014 for shaping the future of information and communications technology. She received the Outstanding service award from the IEEE ComSoc Signal Processing and Communications Electronics technical committee in 2019 and the Outstanding service award from the IEEE ComSoc Women in Communications Engineering Standing Committee in 2020. She received the IEEE ComSoc/KICS Exemplary Global Service Award in 2022. 
\end{IEEEbiography}
\vspace{-1.2cm}

\begin{IEEEbiography}[{\includegraphics[width=1in,height=1.25in,clip,keepaspectratio]{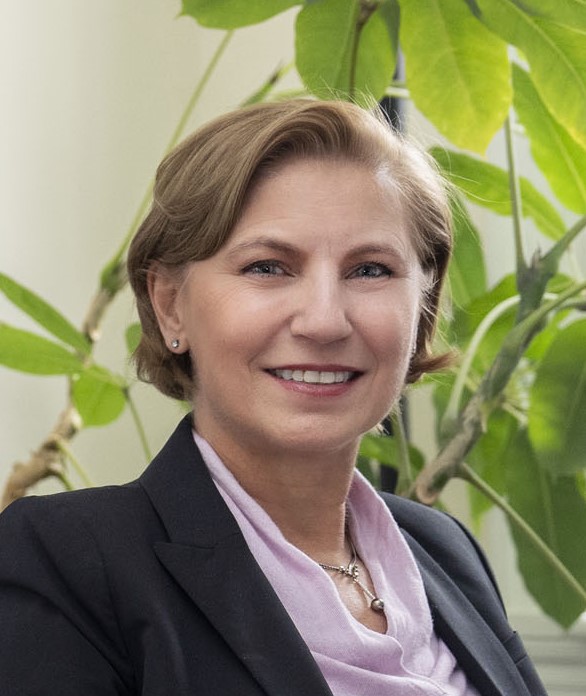}}]{Maïté Brandt-Pearce} (Fellow, IEEE) is a professor of Electrical Engineering and vice provost for faculty affairs at the University of Virginia. She joined UVa after receiving her Ph.D. in Electrical Engineering from Rice University in 1993. Her research interests include free-space optical communications, visible light communications, nonlinear effects in fiber-optics, and cross-layer design of optical networks subject to physical layer degradations.  Dr. Brandt-Pearce is the recipient of an NSF CAREER Award and an NSF RIA. She is a co-recipient of Best Paper Awards at ICC 2006 and GLOBECOM 2012.  She has served on the editorial board of IEEE Transactions of Communications, IEEE Communications Letters, IEEE/OSA Journal of Optical Communications and Networks, IEEE Transactions of Information Theory,  and Springer Photonic Network Communications.    She was Jubilee Professor at Chalmers University, Sweden, in 2014. After serving as General Chair of the Asilomar Conference on Signals, Systems \& Computers in 2009, she served as Technical Vice-Chair of GLOBECOM 2016. She is a member of Tau Beta Pi, Eta Kappa Nu, and a Fellow of the IEEE. In addition to co-editing a book entitled Cross-Layer Design in Optical Networks, Springer Optical Networks Series, 2013, Prof. Brandt-Pearce has over two hundred peer-reviewed technical publications.
\end{IEEEbiography}


\vfill

\end{document}